\documentclass[12pt,a4paper]{article}
\usepackage[T1]{fontenc}
\usepackage{fancyhdr}
\usepackage{amsmath,amsthm,amsfonts,amssymb}
\allowdisplaybreaks% page breaks allowed in align equations
\usepackage{epic,eepic}
\usepackage{graphicx}
\usepackage{authblk}
\usepackage{fullpage}
\usepackage[active]{srcltx}
\usepackage{enumitem}% itemize options
\usepackage{booktabs}% for tables

\usepackage{framed,color}

\usepackage{float}%
\floatstyle{plaintop}%
\restylefloat{table}% these three lines put caption for tables on top

\usepackage[font={small,it}]{caption}% the options makes the font italic in floats
\usepackage[table]{xcolor}% use colours in table rows

\usepackage[toc,page]{appendix}
\graphicspath{{figures/}}

\usepackage[style=authoryear,backend=bibtex,maxcitenames=2,maxbibnames=99]{biblatex}
\bibliography{SLPM}

\usepackage{chngcntr}
\usepackage{dsfont}% indicator function \mathds{1} symbol

\usepackage{algpseudocode}
\usepackage{algorithm}
\usepackage{setspace}% for the pseudocode

\usepackage{hyperref}

\newcommand{\blambda}{\boldsymbol{\lambda}}
\newcommand{\btheta}{\boldsymbol{\theta}}
\newcommand{\bgamma}{\boldsymbol{\gamma}}
\newcommand{\bdelta}{\boldsymbol{\delta}}
\newcommand{\bphi}{\boldsymbol{\phi}}
\newcommand{\bPhi}{\boldsymbol{\Phi}}
\newcommand{\tlambda}{\tilde{\lambda}}
\newcommand{\tblambda}{\tilde{\boldsymbol{\lambda}}}
\newcommand{\talpha}{\tilde{\alpha}}
\newcommand{\tbalpha}{\tilde{\boldsymbol{\alpha}}}
\newcommand{\tbeta}{\tilde{\beta}}
\newcommand{\tbbeta}{\tilde{\boldsymbol{\beta}}}
\newcommand{\tdelta}{\tilde{\delta}}
\newcommand{\tbdelta}{\tilde{\boldsymbol{\delta}}}
\newcommand{\ta}{\tilde{a}}
\newcommand{\tba}{\tilde{\textbf{a}}}
\newcommand{\tb}{\tilde{b}}
\newcommand{\tbb}{\tilde{\textbf{b}}}

\newcommand{\tbphi}{\tilde{\boldsymbol{\phi}}}
\newcommand{\teta}{\tilde{\eta}}
\newcommand{\tzeta}{\tilde{\zeta}}

\makeatletter
\AtEveryBibitem{%
  \global\undef\bbx@lasthash%
  \clearfield{}}
\makeatother

\theoremstyle{plain}

\newtheorem{proposition}[]{Proposition}

\theoremstyle{definition}

\theoremstyle{remark}

\numberwithin{equation}{section}
\numberwithin{figure}{section}

\date{}
\title{\textbf{The Sparse Latent Position Model for nonnegative weighted networks}}
\author{Riccardo Rastelli}
\affil{\footnotesize riccardo.rastelli@wu.ac.at}
\affil{\footnotesize Institute for Statistics and Mathematics, Vienna University of Economics and Business, Austria.}

\begin{document}
\rowcolors{2}{gray!25}{white}% use colours in table's rows
\counterwithout{figure}{section}
\counterwithout{figure}{subsection}
\counterwithout{equation}{section}
\counterwithout{equation}{subsection}

\maketitle
\begin{abstract}
\noindent
This paper introduces a new methodology to analyse bipartite and unipartite networks with nonnegative edge values.
The proposed approach combines and adapts a number of ideas from the literature on latent variable network models.
The resulting framework is a new type of latent position model which exhibits great flexibility, 
and is able to capture important features that are generally exhibited by observed networks, 
such as sparsity and heavy tailed degree distributions.
A crucial advantage of the proposed method is that 
the number of latent dimensions is automatically deduced from the data in one single algorithmic framework.
In addition, the model attaches a weight to each of the latent dimensions, hence providing a measure of their relative importance.
A fast variational Bayesian algorithm is proposed to estimate the parameters of the model.
Finally, applications of the proposed methodology are illustrated on both artificial and real datasets, 
and comparisons with other existing procedures are provided.
\\

\end{abstract}

\baselineskip=20pt
\section{Introduction}\label{sec:introduction}
Statistical models for networks are most often designed for binary interactions between entities.
However, in many cases, the edges carry additional information that may provide crucial insights in the analyses.
One common situation is that of nonnegative interaction values, where each edge carries a positive real number representing, for example, the intensity or the length of the interaction.
This type of scenario is extremely common and it tends to arise in a variety of fields:
in biology, microarray data measure the expression level of genes among a set of samples \parencite{brunet2004metagenes};
gene co-expression networks characterise the similarities between genes based on their expression level \parencite{zhang2005general};
and mutualistic networks describe how often a pollinator visits certain plants \parencite{dormann2008introducing}.
In neuroscience, nonnegative networks may be used to represent functional connectivity levels between different areas of the brain \parencite{bullmore2009complex}.
In finance, the nonnegative values may refer to the liabilities of one financial institution towards others \parencite{gandy2016bayesian}.

The available literature on the analysis of weighted networks, and, in particular, of nonnegative weighted networks, is limited.
This is especially true for the family of Latent Position Models (LPM), which have been widely used in statistical network analyses since the seminal work of \textcite{hoff2002latent}.
In these network models, one postulates that the nodes are points embedded in a latent space, 
and that the propensity to create edges is determined by the Euclidean distance between the two nodes involved.
To the best of my knowledge, in the stream of literature that originated from \textcite{hoff2002latent}, 
the only extension of the original LPM to the analysis of weighted interactions is due to \textcite{sewell2016latent}.

Another fundamental model used in the statistical analyses of networks is the Stochastic Blockmodel (SBM) of \textcite{wang1987}, 
or its equivalent bipartite variant, called Latent Blockmodel (LBM, \cite{govaert2003clustering}).
These two models assume a partition on the set of nodes of the graph, and postulate that edge probabilities are determined by the cluster memberships of the nodes at its extremities.
Extensions of the SBM and LBM to the analysis of weighted networks are also few.
In this context, two relevant papers are those of \textcite{mariadassou2010uncovering} and \textcite{aicher2014learning}.
Recent reviews of the LPM, SBM and their extensions may be found in \textcite{salter2012review,matias2014modeling,raftery2017comment}.

This paper introduces a new modelling approach for the analysis of nonnegative interaction data.
The model proposed, called the Sparse Latent Position Model (SLPM), modifies the original LPM to account for nonnegative weighted edges.
The model adapts a number of ideas from the literature on blockmodels and finite mixture models:
this results in a hybrid framework that exhibits great flexibility, 
and that it is able to elegantly address some of the crucial issues and research questions encountered in LPMs and SBMs.

Similarly to the LPM, the SLPM asserts that nodes are characterised by a $K$-dimensional vector of coordinates.
However, while in the LPM the edge values are determined by the latent Euclidean distances, 
the SLPM postulates that each dimension gives a separate contribution to the edge values,
and that each of these contributions is, in fact, determined by the distance between the two nodes in the corresponding dimension.
Essentially, the SLPM may be interpreted as a finite mixture of $K$ unidimensional LPMs in a weighted network context.
This paper illustrates that the structure of the SLPM allows great flexibility,
and that it can naturally capture relevant features of the the topologies of observed networks, such as sparsity of edges and heavy tailed degree distributions.

In addition, one important facet of the SLPM is the sparsity with respect to the mixture components, 
inherited from recent works on Gaussian finite mixtures \parencite{malsiner2016model,malsiner2017identifying} 
and other types of latent variable models.
In a sparse mixture model, one deliberately creates overfitting by considering a relatively large number of latent groups, 
however, once an appropriate shrinkage prior is defined on the mixing proportions, the superfluous groups are emptied during the estimation.
This approach is strongly motivated by the theoretical results of \textcite{rousseau2011asymptotic}, 
where the authors establish the necessary conditions for the posterior distribution to concentrate around the correct number of groups.

In the SLPM, the sparse representation ensures that the number of mixture components is deduced from the data.
Since these groups correspond to the latent dimensions, the dimensionality of the latent space is automatically estimated from the data.
Furthermore, the mixing proportions of the mixture acquire a rather interesting interpretation, 
since they measure how useful each latent dimension is in explaining the observed data.

Regarding the estimation of the SLPM, a variational Bayesian method \parencite{attias1999inferring} is proposed.
Variational approximations have been largely used in the last two decades in a variety of latent variable models (a recent review may be found in \cite{blei2017variational}).
Although, from a theoretical perspective, the implications of the approximation are not yet completely understood,
a variety of applications \parencite{blei2017variational} have proved that these methods can offer a number of appealing advantages during the estimation process.

Variational approximations have been proposed both for LPM frameworks \parencite{salter2013variational,gollini2016joint} 
and for the SBM and its extensions \parencite{daudin2008mixture,airoldi2008mixed,latouche2010modeles,matias2017statistical,bouveyron2018stochastic}.
In the SLPM context, the variational Bayesian method is rather convenient, since, 
differently from sampling-based approaches, it relies on a fast optimisation framework.
Hence, it allows one to bypass the non-identifiability issues associated with the cluster membership variables and with the latent positions, 
while, at the same time, providing a rough approximation of the posterior uncertainty.
The variational algorithm described in this paper has been implemented in the \texttt{R} package \texttt{SparseLPM}, 
which is publicly available from \texttt{CRAN} \parencite{rcoreteam}.

The paper is organised as follows: in Section \ref{sec:SLPM} the SLPM is formally introduced and studied.
The section contains a detailed description of its main properties, and a number of comparisons with other well known models and with the existing literature.
Section \ref{sec:inference} describes the variational algorithm, and the details regarding the optimisation procedure, including a useful initialisation strategy.
Finally Sections \ref{sec:simulations} and \ref{sec:applications} illustrate applications of the methodology to artificial and real datasets, 
respectively.

\section{The Sparse Latent Position Model}\label{sec:SLPM}
\subsection{The model}\label{sec:SLPM_model}
The observed data matrix $\textbf{X}$ has $M$ rows and $N$ columns, and its entries are nonnegative.
In a network context, $\textbf{X}$ defines the nonnegative adjacency matrix of a bipartite graph, 
whereby the nodes are separated into two sets (of cardinalities $M$ and $N$, respectively), and only the interactions between nodes of different types are allowed. 
It is important to note that, as a special case, 
the unipartite weighted graph structure is recovered if $\textbf{X}$ is a square matrix and its rows and columns are indexed with labels from the same set.

The observations $\textbf{X} = \left\{x_{ij}: i=1,\dots,M;\ j = 1,\dots,N\right\}$ are assumed to be independent realisations from a finite mixture of $K$ exponential densities:
\begin{equation}\label{eq:slpm_mixture_1}
p\left(x_{ij}\middle\vert \blambda, \btheta_{ij}\right) = \sum_{k=1}^{K} \lambda_k f_{\mathcal{E}}\left(x_{ij}; \theta_{ijk}\right)
\end{equation}
where $f_{\mathcal{E}}\left( \cdot;\theta \right)$ denotes an exponential density with rate $\theta$, 
and the vector $\btheta_{ij} = \left\{\theta_{ij1},\dots,\theta_{ijK}\right\}$ indicates the component-specific rates, for every edge $(i,j)$.

In the SLPM, the $K$ groups of the mixture distribution are re-interpreted as the $K$ latent dimensions of a LPM, 
where each dimension gives a separate contribution in determining the probabilities of the edges.
In particular, the latent variable $\theta_{ijk}$ corresponds to the squared Euclidean distance between $i$ and $j$ in the $k$-th dimension, i.e.:
\begin{equation}\label{eq:slpm_theta_1}
\theta_{ijk} = \left(U_{ik} - V_{jk}\right)^2
\end{equation}
where $U_{ik}$ (resp. $V_{jk}$) indicates the $k$-th coordinate of $i$ (resp. $j$) in the latent space $\mathbb{R}^K$.
Since the expectation of an exponential random variable is equal to its inverse rate, 
\eqref{eq:slpm_theta_1} guarantees that the generated edge weight is expected to be large (resp. small) if the two nodes are close (resp. far apart) in every latent dimension.

As is common in mixture modelling, an equivalent representation of the same model can be obtained using data augmentation.
An allocation variable $Z_{ij}$ is attached to the edge $(i,j)$, for all $1\leq i \leq M$ and $1\leq j \leq N$, such that:
\begin{equation}
\rowcolors{1}{}{}
Z_{ij} = \begin{cases}
	  1 & \mbox{ with probability } \lambda_1\\
	  \vdots & \hspace{1cm}\vdots\\
	  K & \mbox{ with probability } \lambda_K
	  \end{cases}
\end{equation}
Now, the conditional log-likelihood reads as follows:
\begin{equation}\label{slpm_likelihood_1}
\ell_{\textbf{X}}\left(\mathcal{Z},\textbf{U},\textbf{V}\right) = 
\sum_{i=1}^{M} \sum_{j=1}^{N} \sum_{k=1}^{K} Z_{ijk}\left\{\log\left[\left(U_{ik} - V_{jk}\right)^2\right] - x_{ij}\left(U_{ik} - V_{jk}\right)^2\right\}
\end{equation}
where $Z_{ijk} = 1$ if $Z_{ij}= k$ and $Z_{ijk} = 0$ otherwise.

Finally, additional hierarchical levels are considered for the mixing proportions and for the latent positions:
\begin{equation}
\begin{split}
\left( \lambda_1,\dots,\lambda_K \right) &\sim Dirichlet\left(\delta_1,\dots,\delta_K\right)\\
U_{ik} &\sim Gaussian\left(0,1/\gamma_k\right)\\
V_{jk} &\sim Gaussian\left(0,1/\gamma_k\right)\\
\gamma_k &\sim Gamma\left(a_k,b_k\right)
\end{split}
\end{equation}
for all $1\leq i \leq M$, $1\leq j \leq N$ and $1\leq k \leq K$. 
Here, $Gaussian$ refers to the univariate normal density.

\begin{figure}[htb]
 \centering
 \includegraphics[width = 0.5\textwidth]{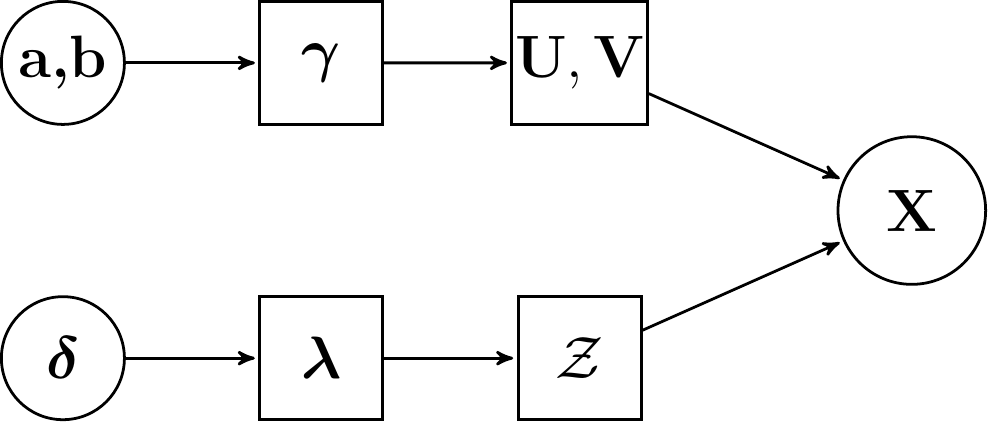}
 \caption{Graphical summary of the dependencies in the SLPM.}
 \label{fig:graphical_model}
\end{figure}

The dependencies between the model parameters are summarised in the graphical representation of Figure \ref{fig:graphical_model}.

\subsection{Interpretation and properties}\label{sec:SLPM_properties}

\paragraph{A flexible mixture model.} 
Mixtures of exponential distributions satisfy a number of interesting properties, and have been well studied and widely applied in a variety of fields.
The reader may find an overview of these properties in \textcite{fruhwirth2006finite} and references therein.
In particular, one key property of these models is their flexibility: 
Bernstein's theorem states that any completely monotone probability density function may be written as a continuous mixture of exponentials (\cite{feldmann1998fitting}, Theorem 3.1). 
In the present paper, this argument motivates the structure proposed in \eqref{eq:slpm_mixture_1}, 
asserting that the SLPM may be particularly appropriate to represent extremely complex network structures, when $K$ is large enough.

\begin{figure}[htbp]
\centering
\includegraphics[width=0.49\textwidth]{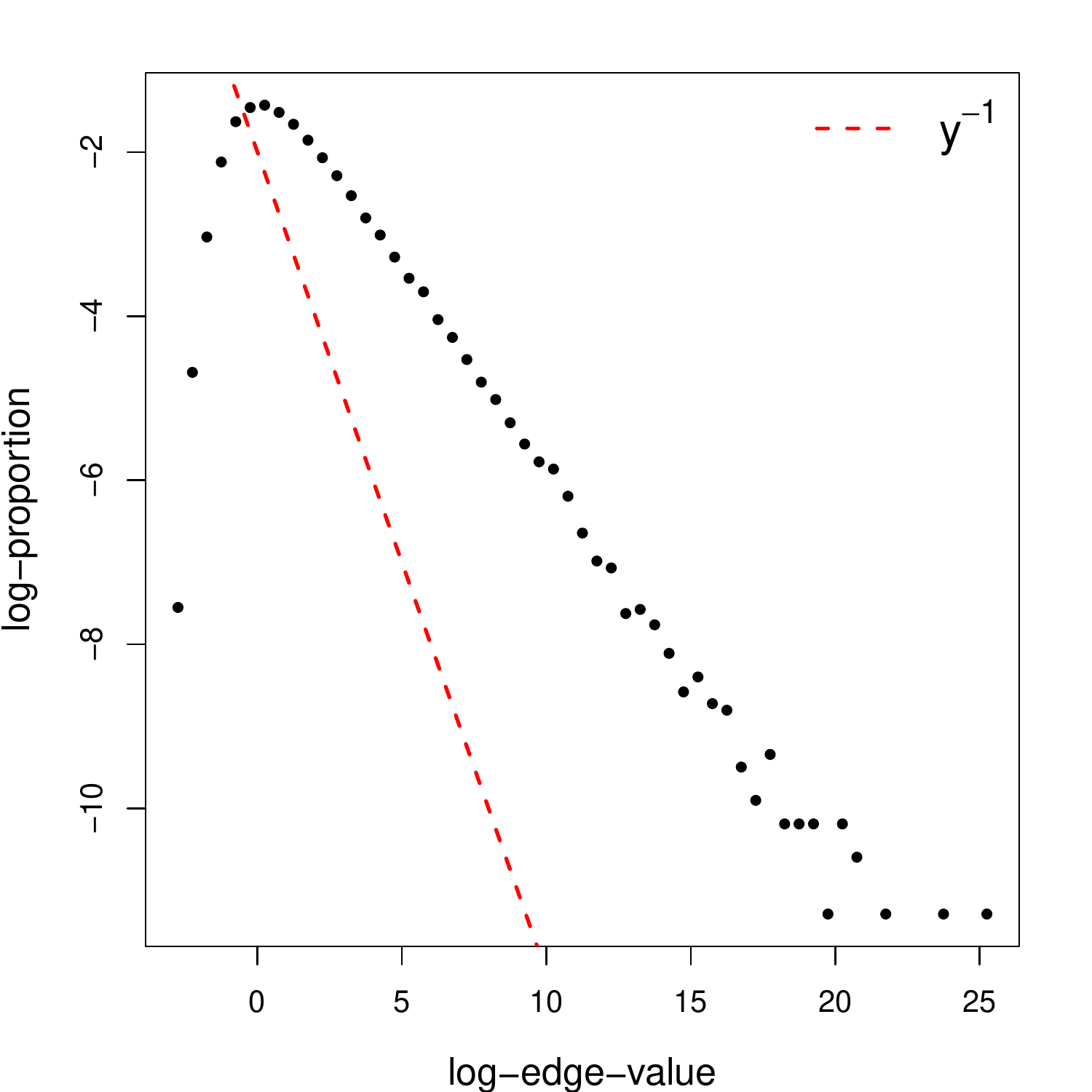}
\includegraphics[width=0.49\textwidth]{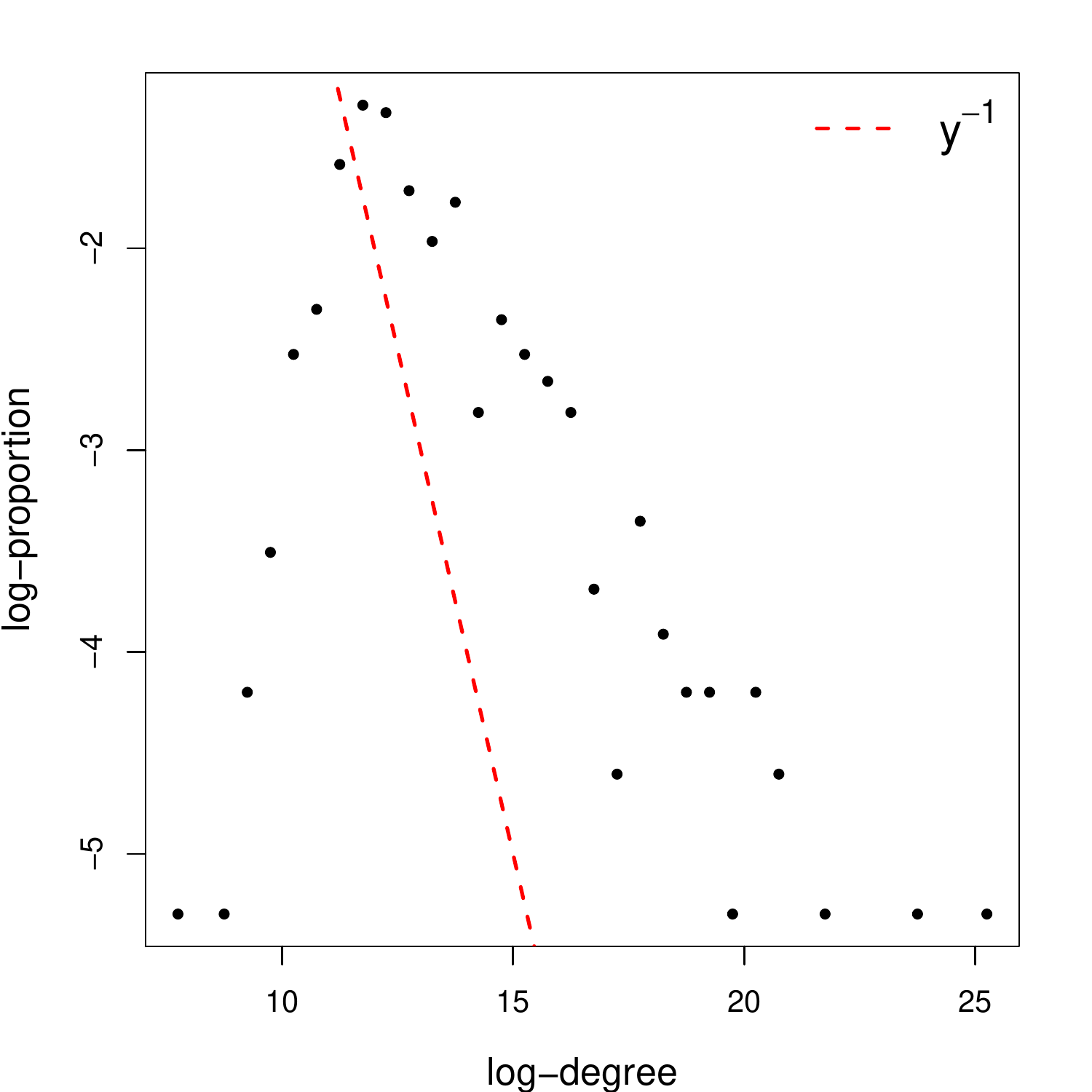}
\caption{Log-log scale empirical distributions for the edge weights (left panel) and out-degrees (right panel) of a simulated SLPM with $400$ nodes and $2$ latent dimensions.}
 \label{fig:sim_dd_12}
\end{figure}

\paragraph{Weight and degree distributions.} 
The SLPM is able to capture relevant features that are usually exhibited by observed networks. 
Figure \ref{fig:sim_dd_12} uses simulated data to demonstrate that the SLPM can naturally represent highly skewed and extremely heavy tailed weight and degree distributions.
In fact, both panels show that the tails of the weight and degree distributions decay to zero slower than a power-law with exponent $1$.
This is a desirable feature in many types of applications: 
for example, the topologies of social networks and financial networks are known to be characterised by the presence of hubs and by core-periphery structures 
\parencite{barabasi1999emergence,newman2001structure,boss2004network}.

\paragraph{Sparsity.}
In recent years, sparse finite mixtures have received a lot of attention (see \cite{malsiner2016model} and references therein) 
since they are able to mimic the properties displayed by other common models
(for example by the Dirichlet process mixture of \cite{quintana2003bayesian}, as shown in \cite{Fruehwirth-Schnatter2018}), 
while maintaining a very simple computational framework to perform inference.
More importantly, sparse finite mixtures are able to efficiently assess many competing models in one single run of the estimation algorithm.
In fact, they do not require grid searches over all the possible values of $K$, 
nor trans-dimensional samplers such as the reversible jump of \textcite{richardson:green97}. 
This generally leads to substantial gains in terms of computational efficiency.

A sparse finite mixture is a deliberately overfitted clustering model where one specifies a shrinkage prior on the mixture weights and on the component-specific parameters 
to ensure that superfluous groups are emptied \parencite{malsiner2016model,malsiner2017identifying}. 
The SLPM adapts this idea in a weighted latent position network model context: 
overfitting is achieved by considering a large number of mixture components, 
while imposing a sparse Dirichlet prior distribution on the mixing proportions to ensure that superfluous groups become empty.
In practice, such sparse Dirichlet prior is attained through $\delta_1 = \dots = \delta_K = \delta < 1/K$, as illustrated by \textcite{rousseau2011asymptotic}.

As a consequence, the SLPM is a type of LPM where a large number of latent dimensions can be considered, hence guaranteeing the theoretical flexibility argued earlier in this section.
At the same time, parsimony is attained using the shrinkage prior on the mixing proportions, making sure that only the relevant latent dimensions are actually used. 
The final product is a LPM where not only the number of latent dimensions is estimated from the data, but also where the dimensions are automatically ranked according to their importance, 
as indicated by the mixing proportions.
This feature becomes very convenient when plotting the latent space: for example, one can select and show only the two most relevant latent dimensions, 
or the two latent dimensions that characterise the large edge values.

Another aspect of the sparsity regards the edge-values: 
most observed weighted networks exhibit zero-inflated empirical distributions, where the majority of edges are absent or carry a null weight.
The SLPM can account for this type of sparsity in a very natural way, 
as edge weights are shrunk towards zero when generated by latent dimensions which are characterised by a small precision $\gamma_k$, for $1\leq k \leq K$.
In practice, in a fitted SLPM, one or more latent dimensions may be exploited to separate the null weights from the rest, hence obtaining a spike-and-slab type of sparsity.

\paragraph{Outliers.}
In the data augmented representation of the SLPM, each edge is assigned to and characterised by one dimension of the latent space.
It is possible that only one edge is assigned to a certain latent dimension.
Since the shrinkage Dirichlet prior heavily penalises this type of behaviour, 
the only circumstance where a fitted SLPM may exhibit this feature is when the value carried by such edge is an outlier that cannot be explained through the other relevant dimensions.
That is, the SLPM may exploit the empty components to separate those edges that cannot be represented by the dimensions that are being used.
This mechanism naturally highlights outliers in the data, or, equivalently, it may signal that, to some extent, the latent space representation is not particularly appropriate for the dataset considered. 

\paragraph{Identifiability.}
The likelihood of a latent position network model is known to be invariant under rotations, translations and reflections of the latent positions \parencite{hoff2002latent}.
Markov chain Monte Carlo techniques may be used to obtain a sample from the posterior distribution of these models, 
however, post-processing becomes then necessary to deal with the identifiability issues and, thus, to attain meaningful interpretations.

The SLPM is no exception in this regard, as its likelihood is invariant under translations and reflections.
In addition, the likelihood is also unaffected by any relabeling of the latent dimensions, which, in turn, leads to the same label switching issues studied by \parencite{stephens2000dealing}.
The combination of these two sources of non-identifiability makes the available post-processing methods inadequate, and discourage the use of sampling-based methods, as their output cannot be efficiently interpreted.
This paper considers an optimisation framework which bypasses all identifiability problems and makes the label switching and the translation invariance irrelevant.

\paragraph{Univariate networks.}
Section \ref{sec:SLPM_model} introduces the SLPM as a statistical model for bipartite networks. 
The canonical directed unipartite framework is obtained as a special case, when $M=N$ and the nodes' labels vary in the same set. 
The SLPM can be easily modified to neglect self-edges, if these are not of interest.

If a directed unipartite scenario is considered, the SLPM separates the latent positions of nodes based on whether they are sending or receiving an edge.
For example, the latent position of node $i$ is given by $\textbf{U}_i = \left\{ U_{i1},\dots,U_{iK} \right\}$ when sending edges 
and by $\textbf{V}_i = \left\{ V_{i1},\dots,V_{iK} \right\}$ when receiving them.
Although unusual, such a distinction may give interesting advantages when disassortative patterns\footnote{Disassortative 
mixing refers to the tendency of nodes to connect to those that are not similar to them. 
For example, disassortative mixing arises when many nodes are connected only to a large hub (core-periphery structure).
In the SLPM context, the distinction between sender and receiver position may be useful to capture the fact that 
certain nodes act as bridges, receiving edges from a set of nodes and sending edges into a different set.} 
are observed.

Finally, with some additional work, it may also be possible to extend the SLPM framework to handle undirected unipartite networks, 
where the latent positions $\textbf{U}$ and $\textbf{V}$ coincide.
However, this scenario is not explored in the present paper and it is left to future work.

\subsection{Comparisons with other models}\label{sec:SLPM_comparisons}

\paragraph{Latent position models.}

\textcite{rastelli2016properties} illustrate the properties exhibited by the basic LPM, 
and show that this model can become very flexible when additional nodal random effects are added.
In particular, they show that the basic LPM and its extensions are naturally designed to capture transitivity, homophily and heavy tailed degree distributions.
In the SLPM there are no nodal random effects, and, similarly to the LPM, each of the nodes in the graph is characterised by a number of latent coordinates.
However, the generative mechanism behind the SLPM makes use of these quantities in a different way, 
as each edge is determined by the coordinates in only one of the latent dimensions.

On the one hand, this modification makes the model much more flexible: 
for example, differently from the canonical LPM, two nodes located in the same position may not be stochastically equivalent conditionally on the allocation variables, 
in that the distributions generating their edges may be completely different.
Lack of transitivity, which is common in unipartite networks with strong core-periphery structures, may also be represented by the SLPM:
conditionally on $\mathcal{Z}$, the variables $X_{ij}$ and $X_{i'j}$ are dependent only if they are allocated to the same latent group, for any $i,i',j$.
In addition, the SLPM can capture homophily and, to some extent, the lack of it.
In fact, two nodes that are close in the latent space are expected to be strongly connected, regardless of which dimension determines the edge weight.
Nevertheless, if two nodes are close in at least one dimension and far apart in at least another, a wide range of scenarios become possible.

On the other hand, the SLPM is less interpretable than the LPM and its extensions.
In fact, in the SLPM, the Euclidean distance between two $K$ dimensional coordinates vectors simply has no relevant meaning, 
as one should rather study the distances between nodes for each latent dimension separately.
In addition, a fitted SLPM may be characterised by a large number of latent dimensions, making graphical representations impractical.
So, generally speaking, extracting meaningful interpretations from a two-dimensional representation of the latent space of a SLPM
may not be as straightforward as in other LPM frameworks.

One significant novelty introduced by the SLPM is that the number of latent dimensions is estimated from the data, and, in addition, a weight is assigned to each dimension to represent its importance.
The estimation of $K$ is a long standing problem that very few times has been directly addressed in the LPM literature;
the main reason being that the arbitrary choice $K=2$ provides a convenient framework to clearly show the results.
One relevant work in this context is \textcite{handcock2007model}, where the authors introduce a variant of the well-known Bayesian Information Criterion (BIC) to perform model choice.
However, in the discussion they also point out that this method may not be formally correct and appropriate when choosing the number of latent dimensions.
Another work is that of \textcite{friel2016interlocking}, where the authors use instead the Deviance Information Criterion (DIC).
One drawback of this approach is that it requires the Markov chain Monte Carlo sampler to be run for every value of $K$ considered.
By contrast, in the SLPM, the model-choice problem is addressed in one single algorithmic framework, since the superfluous groups are automatically emptied during the estimation.

Another crucial difference with the existing literature relates to the presence of weights on the edges: the aforementioned LPMs can only deal with binary edges. 
The only paper that proposes an adaptation of the LPM framework to a weighted network context is \textcite{sewell2016latent}.
The authors consider a dynamic weighted network scenario, and propose several types of models based on the type of weights carried by the edges.
For the nonnegative case, they propose a Tobit model \parencite{tobin1958estimation}, and estimate model parameters using Markov chain Monte Carlo sampling.
The SLPM is fundamentally different, as it uses a mixture model to characterise the distribution of the edges, 
with the goal of naturally representing the sparsity of network connections, and estimating the number of latent dimensions.
While the model of \textcite{sewell2016latent} may be better suited to clearly summarise the network data in a two-dimensional space, 
dynamic extensions of the proposed SLPM should provide a more flexible framework to predict missing or future values.

\paragraph{Weighted blockmodels.}
The stochastic blockmodel (SBM) of \textcite{wang1987} has been extended to account for weighted interactions by \textcite{mariadassou2010uncovering} and by \textcite{aicher2014learning}.
In the SBM, the nodes are characterised by a latent cluster membership, and any two nodes assigned to the same group are said stochastically equivalent, 
as their connections are generated from the same parametric distribution.
Equivalently, one may say that each edge of the network is characterised by a latent cluster membership, which is itself determined by the allocations of the nodes at its extremities:
the edges assigned to the same group are generated from the same parametric distribution.
This mechanism also characterises the well-known mixed-membership stochastic blockmodel (MMSBM) of \textcite{airoldi2008mixed}, 
where, again, the cluster membership characterising each edge is determined by node-specific attributes.
One crucial advantage of the MMSBM with respect to the SBM is that it allows heterogeneity to be present within any subset of nodes.

The SLPM may be seen as a variant of the SBM and the MMSBM, where the cluster membership assigned to an edge is not determined by nodal information, but rather by some global parameters 
(the mixing proportions $\blambda$).
In addition, once these cluster memberships are set, the distributions generating the edge weights are chosen not only according to such clustering variables, 
but also using cluster-specific nodal information, i.e. the latent coordinates.
Hence, the SLPM relates to the literature originated from \textcite{airoldi2008mixed} 
and it extends the classic blockmodels by introducing an additional ingredient, the latent distance representation, 
to explain the observed heterogeneity and to increase the flexibility of the mixture.

\paragraph{Tensor factorisation models.}
Tensors are array objects that generalise adjacency matrices to more than two dimensions, 
and are generally used to represent multiview networks, dynamic networks, and other types of multivariate interaction data.
One common approach used to model these objects is inspired by factor models \parencite{bartholomew2011latent}, 
where the tensor (or, in our context, the adjacency matrix $\textbf{X}$) is decomposed using a low-rank factorisation:
\begin{equation}
 \textbf{X} = \textbf{UV}' + \textbf{E}
\end{equation}
Here, $\textbf{U}$ has size $M\times K$, $\textbf{V}$ has size $N\times K$, $\textbf{E}$ has size $M\times N$, and the dash denotes matrix transposition.
A very rich machine learning literature is available on this topic (see, for example, \cite{lim2007variational,mnih2008probabilistic,salakhutdinov2008bayesian}), 
however, since this paper deals with nonnegative matrices, 
our interest mainly focuses on the nonnegative matrix factorisation (NMF) approach of \textcite{lee1999learning,lee2001algorithms}.

In statistics, similar models have been considered for the analysis of binary networks by \textcite{hoff2002latent}, under the name of \textit{projection models}.
Since each edge is assumed to be generated through a transformation of the scalar product of $K$-dimensional latent vectors, 
the latent space may be interpreted as a $K$-dimensional sphere, where a small angle between two nodes implies a higher probability of connection.
While this generative mechanism shares similarities with LPMs and other distance based models, 
it provides a vastly different geometrical representation of the network, hence, it may lead to qualitatively different statistical analyses and interpretations.

Projection models have been intensely studied and extended in a number of ways \parencite{hoff2005bilinear,durante2014nonparametric}, 
and have also been considered as models for weighted interactions \parencite{hoff2018additive}.
The work of \textcite{durante2017nonparametric} has several similarities with the approach proposed in this paper.
The authors consider a multiview network and propose a mixture of projection models to explain its binary edges.
They exploit the sparse Dirichlet prior on the mixture weights to ensure that non-relevant mixture components are emptied.
They propose a Markov chain Monte Carlo sampler and focus their analysis only on the characterisation of the edge probabilities, hence bypassing the noted identifiability issues.

The SLPM relies instead on a sparse finite mixture representation of the latent space within a distance-based framework, which also leads to an estimation of $K$.
The new model proposed in this paper is specifically designed for nonnegative weighted bipartite networks, and, hence, it could be seen as a competitor of the NMF algorithms.
Here, the main novelty introduced by the SLPM with respect to NMF and the other tensor factorisation models is the latent distance framework, 
where interaction values are determined by how \textit{close} (in the Euclidean sense) the two entities are in the latent space.

\section{Inference}\label{sec:inference}
According to Section \ref{sec:SLPM_model} and Figure \ref{fig:graphical_model}, the posterior density of a SLPM factorises as follows:
\begin{equation}\label{eq:slpm_posterior_1}
\pi\left( \mathcal{Z},\textbf{U},\textbf{V},\blambda,\bgamma \middle\vert \textbf{X}, \bdelta, \textbf{a}, \textbf{b}\right) 
\propto \mathcal{L}_{\textbf{X}}\left(\mathcal{Z},\textbf{U},\textbf{V}\right) 
\pi\left( \mathcal{Z}\middle\vert \blambda \right) \pi\left( \blambda\middle\vert\bdelta \right)
\pi\left( \textbf{U}\middle\vert\bgamma \right)\pi\left( \textbf{V}\middle\vert\bgamma \right) \pi\left( \bgamma\middle\vert\textbf{a},\textbf{b} \right)
\end{equation}
However, as noted in the previous section, posterior samples for the model parameters are not interpretable due to the multiple sources of non-identifiabilities.
This means that a traditional Markov chain Monte Carlo sampler cannot be used to characterise the posterior and, hence, the uncertainty around the parameter estimates.
To overcome this limitation, an approach based on Variational Bayes (VB, \cite{attias1999inferring,attias2000variational}) is proposed.
This means that sampling techniques are replaced by optimisation, which allows much faster computations while still providing a reasonable characterisation of the posterior uncertainty.

\subsection{Variational Bayes}
In this paper, the theory behind VB is not discussed in detail: 
the interested reader may find more complete illustrations of these methods for example in \textcite{latouche2010modeles,blei2017variational} and references therein.

Denote the collection of model parameters $\left\{ \mathcal{Z},\textbf{U},\textbf{V},\blambda,\bgamma \right\} = \bphi\in\bPhi$: 
the goal of VB is to perform a search in a family $\mathcal{Q}$ of densities over $\bPhi$, to find the density $q^*(\cdot)\in\mathcal{Q}$ that is closest, 
in the sense of Kullback-Leibler (KL) divergence\footnote{The KL divergence measures the discrepancy between two densities and it is defined as
$$
KL\left( q\middle\vert\middle\vert p \right) = \mathbb{E}_q\left[ \log q\left( \bphi \right)\right] - \mathbb{E}_q\left[ \log p\left( \bphi \right)\right]
$$
Note that the KL is not a distance since it is not symmetric, 
i.e. $KL\left( q\middle\vert\middle\vert p \right) \neq KL\left( p\middle\vert\middle\vert q \right)$,
however, it is nonnegative and it becomes zero when $q$ and $p$ coincide.}, to the posterior density $\pi\left( \bphi\middle\vert \textbf{X} \right)$ defined by \eqref{eq:slpm_posterior_1}.

Generally, the densities in the family $\mathcal{Q}$ are, by construction, smoother, more tractable, and easier to interpret than $\pi\left( \cdot\middle\vert \textbf{X} \right)$.
As is common practice, the variational densities in $\mathcal{Q}$ are assumed to satisfy a mean-field assumption:
\begin{equation}\label{eq:vb_mean_field}
\begin{split}
 q\left( \mathcal{Z},\textbf{U},\textbf{V},\blambda,\bgamma \right) = &
 \left\{ \prod_{i,j}q_{\mathcal{Z}}\left( Z_{ij} \right)\right\}
 \left\{ \prod_{i,k}q_{\textbf{U}}\left( U_{ik} \right)\right\}
 \left\{ \prod_{j,k}q_{\textbf{V}}\left( V_{jk} \right)\right\} \\
 &\times\left\{ \prod_{i,j}q_{\blambda}\left( \blambda_{ij} \right)\right\}
 \left\{ \prod_{k}q_{\bgamma}\left( \gamma_{k} \right)\right\}
\end{split}
\end{equation}
In \eqref{eq:vb_mean_field} and from here onwards: $1\leq i\leq M$, $1\leq j\leq N$ and $1\leq k\leq K$, unless otherwise specified.
Similarly to most other LPMs, the SLPM does not have all of its full-conditional distributions in standard form, mainly due to the likelihood specification. 
In a MCMC context, this means that non-standard parameter updates must be performed using Metropolis-within-Gibbs \parencite{hoff2002latent}.
In a VB context, this implies that the parametric forms of the variational densities in $\mathcal{Q}$ cannot be automatically derived, 
and that closed forms updates will not be available for the variational parameters.

Hence, the following variational densities are assumed:
\begin{equation}
 \begin{split}
  q_{\mathcal{Z}}\left( Z_{ij1},\dots,Z_{ijK} \middle\vert \tlambda_{ij1},\dots,\tlambda_{ijK}\right) &\sim Multinomial\left( 1; \tlambda_{ij1},\dots,\tlambda_{ijK}\right) \\
  q_{\textbf{U}}\left( U_{ik} \middle\vert \talpha_{Uik}, \tbeta_{Uik}\right) &\sim Gaussian\left( \talpha_{Uik}, \tbeta_{Uik}\right) \\
  q_{\textbf{V}}\left( V_{jk} \middle\vert \talpha_{Vjk}, \tbeta_{Vjk}\right) &\sim Gaussian\left( \talpha_{Vjk}, \tbeta_{Vjk}\right) \\
  q_{\blambda}\left( \lambda_{ij1},\dots,\lambda_{ijK} \right) &\sim Dirichlet\left( \tdelta_{1}, \dots, \tdelta_{K}\right) \\
  q_{\bgamma}\left( \gamma_{k} \right) &\sim Gamma\left( \ta_k, \tb_k\right)
 \end{split}
\end{equation}
Here, $Gaussian$ refers to the univariate normal density.
It should be noted that all the variational parameters are indicated with a tilde: 
$$
\tbphi = \left\{ \tblambda, \tbalpha_{U}, \tbbeta_{U}, \tbalpha_{V}, \tbbeta_{V}, \tbdelta, \tba, \tbb \right\}
$$

\begin{proposition}\label{prop:elbo}
 Minimising the KL divergence between $q\left( \cdot \right) \in \mathcal{Q}$ and the SLPM posterior $\pi\left( \cdot\middle\vert\textbf{X} \right)$ 
 is (approximately) equivalent to maximising the following functional with respect to the variational parameters $\tbphi$:
 \small{
 \begin{equation}\label{eq:elbo_1}
 \begin{split} 
  \mathcal{F}\left( \tbphi \right) = &
  \mbox{ const } + \sum_{i,j,k} \tlambda_{ijk}\left[ \psi\left( \frac{\teta_{ijk}^2}{\tzeta_{ijk}} \right) - \log\left( \teta_{ijk} \right) + \log\left( \tzeta_{ijk} \right) - x_{ij}\teta_{ijk} - \log\left( \tlambda_{ijk} \right)\right] \\
  & + \sum_{k} \left( \delta_k-\tdelta_k+\sum_{i,j}\tlambda_{ijk} \right)\left[ \psi\left( \tdelta_k \right) - \psi\left( \sum_h \tdelta_h \right)\right] \\
  & + \sum_k \left( a_k - \ta_k + \frac{M+N}{2} \right) \left[ \psi\left( \ta_k \right) -\log\tb_k \right] 
  - \sum_{k}\frac{\ta_k}{\tb_k}\left( b_k + \frac{\tilde{S}_k}{2} \right)\\
  & + \frac{1}{2}\sum_{ik}\log\left( \tbeta_{Uik} \right) + \frac{1}{2}\sum_{j,k}\log\left( \tbeta_{Vjk} \right) \\
  & - \log\Gamma\left( \sum_k\tdelta_k \right) + \sum_k\left[ \log\Gamma\left( \tdelta_k \right) +  \ta_k - \ta_k\log\tb_k + \log\Gamma\left( \ta_k \right) \right]
 \end{split}
 \end{equation}
 }
 where:
 \begin{equation}
 \begin{split} 
  \teta_{ijk} &= \tbeta_{Uik} + \tbeta_{Vjk} + \left( \talpha_{Uik} - \talpha_{Vjk} \right)^2 \\
  \tzeta_{ijk} &= 2\teta_{ijk} - 2\left( \talpha_{Uik} - \talpha_{Vjk} \right)^4 \\
  \tilde{S}_k &= \sum_i\left( \tbeta_{Uik} + \talpha_{Uik}^2 \right) + \sum_j\left( \tbeta_{Vjk} + \talpha_{Vjk}^2 \right)
 \end{split}
\end{equation}
and $\psi\left( \cdot \right)$ indicates the first derivative of the log-gamma function.
\end{proposition}
The functional $\mathcal{F}$ is often called the variational free energy.
The proof of Proposition \ref{prop:elbo}, and a description of the approximations introduced to obtain this result are provided in Appendix \ref{app:elbo}.

\subsection{Maximisation of \texorpdfstring{$\mathcal{F}$}{Fcal}}
The maximisation of \eqref{eq:elbo_1} is performed using a coordinate ascent algorithm (CAVI, \cite{bishop2006pattern,blei2017variational}),
where each of the variational densities composing $q\left( \cdot \right)$ are updated in turn in a greedy fashion.
This section shows that some of these updates are available in closed form, whereas iterative procedures must be used for the remaining densities.
The proofs are all shown in the appendix.

\begin{proposition}\label{prop:update_lambda}
 For any given configuration of variational parameters $\tbphi$ and indexes $1\leq i \leq M$ and $1\leq j \leq N$, 
 the values of $\left( \tlambda_{ij1},\dots,\tlambda_{ijK} \right)$ that maximise $\mathcal{F}$ are given by:
 \begin{equation}\label{eq:update_lambda_1}
  \tlambda^*_{ijk} = \frac
  {\exp\left\{   \psi\left( \frac{\teta_{ijk}^2}{\tzeta_{ijk}} \right) - \log\left( \frac{\teta_{ijk}}{\tzeta_{ijk}} \right) - x_{ij}\teta_{ijk} + \psi\left( \tdelta_k \right) - \psi\left( \sum_k \tdelta_k \right)   \right\}}
  {\sum_{k'} \exp\left\{   \psi\left( \frac{\teta_{ijk'}^2}{\tzeta_{ijk'}} \right) - \log\left( \frac{\teta_{ijk'}}{\tzeta_{ijk'}} \right) - x_{ij}\teta_{ijk'} + \psi\left( \tdelta_k \right) - \psi\left( \sum_{k''} \tdelta_{k''} \right)   \right\}}
 \end{equation}
for $1\leq k,k',k'' \leq K$.
\end{proposition}

\begin{proposition}\label{prop:update_delta}
 For any given configuration of variational parameters $\tbphi$, the values of $\tbdelta$ that maximise $\mathcal{F}$ are given by:
 \begin{equation}\label{eq:update_delta_1}
  \tdelta^*_{k} = \delta_k + \sum_{i,j} \tlambda_{ijk}
 \end{equation}
for $1\leq k \leq K$.
\end{proposition}

\begin{proposition}\label{prop:update_ab}
 For any given configuration of variational parameters $\tbphi$, the values of $\tba$ and $\tbb$ that maximise $\mathcal{F}$ are given by:
 \begin{equation}\label{eq:update_ab_1}
 \begin{split}
  \ta^*_{k} &= a_k + \frac{M+N}{2} \\
  \tb^*_{k} &= b_k + \frac{1}{2} \sum_i\left( \tbeta_{Uik} + \talpha_{Uik}^2 \right) + \frac{1}{2} \sum_j\left( \tbeta_{Vjk} + \talpha_{Vjk}^2 \right)
 \end{split}
 \end{equation}
for $1\leq k \leq K$.
\end{proposition}

Similar closed form updates are not available for the remaining variational densities $q_{\textbf{U}}$ and $q_{\textbf{V}}$.
Hence, the updates for the corresponding parameters aim at improving the value of $\mathcal{F}$ using a greedy rule in the direction of the natural gradient \parencite{amari1998natural}.

\begin{proposition}\label{prop:update_alpha_beta}
 The update defined by:
 \begin{equation}\label{eq:update_alpha_beta_1}
 \begin{split}
  \talpha^*_{Uik} &= \talpha_{Uik} + \varepsilon\tbeta_{Uik} \frac{\partial \mathcal{F}\left( \tbphi \right)}{\partial \talpha_{Uik}} \\
  \tbeta^*_{Uik} &= \exp\left\{ 2\varepsilon\tbeta_{Uik}\frac{\partial \mathcal{F}\left( \tbphi \right)}{\partial \tbeta_{Uik}}\right\} \tbeta_{Uik}
 \end{split}
 \end{equation}
 is in the direction of the natural gradient of $\mathcal{F}$ with respect to $\talpha_{Uik}$ and $\tbeta_{Uik}$, and, for a small enough $\varepsilon>0$, it does not decrease the value of $\mathcal{F}$.
\end{proposition}
A result equivalent to Proposition \ref{prop:update_alpha_beta} is also provable for the variational density associated to the latent position $V_{jk}$, for every $1 \leq j \leq N$ and $1 \leq k \leq K$.

In practice, the algorithm keeps track of the learning rate $\varepsilon$ associated to the variational densities of each latent position.
To update one of these densities, an initial learning rate equal to $2\varepsilon$ is considered, and new proposed parameter values are calculated according to \eqref{eq:update_alpha_beta_1}.
If the new value of $\mathcal{F}$ is not smaller, the new values are retained and the algorithm proceeds to the following update.
Otherwise, the learning rate is repeatedly halved until it is sufficiently small to guarantee a non-decrease of $\mathcal{F}$.
The pseudocode for the optimisation routine is shown in Algorithm \ref{algorithm_1}.
\begin{algorithm}[htb]
\begin{spacing}{1.2}
\caption{\texttt{CAVI algorithm for the SLPM}}
\label{algorithm_1}
\begin{algorithmic}
\State INPUT: $\texttt{tol}$ and starting values $\tbphi$
\State set $\mathcal{F}^* = \mathcal{F}\left( \tbphi \right)$
\State set $\texttt{stop} = \texttt{false}$
\While{$!\texttt{stop}$}
\State for every $i$ and $j$: update $\tblambda_{ij}$ using \eqref{eq:update_lambda_1}
\State update $\tbdelta$ using \eqref{eq:update_delta_1}
\State update $\tba$ and $\tbb$ using \eqref{eq:update_ab_1}
\State for every $i$ and $k$: update $\talpha_{Uik}$ and $\tbeta_{Uik}$ using \eqref{eq:update_alpha_beta_1}
\State for every $j$ and $k$: update $\talpha_{Vjk}$ and $\tbeta_{Vjk}$ using \eqref{eq:update_alpha_beta_1}
\If{$\mathcal{F}\left( \tbphi \right) \leq \mathcal{F}^* + \texttt{tol}$} $\texttt{stop} = \texttt{true}$
\EndIf
\State set $\mathcal{F}^* = \mathcal{F}\left( \tbphi \right)$
\EndWhile
\State OUTPUT: $\mathcal{F}^*$ and $\tbphi$
\end{algorithmic}
\end{spacing}
\end{algorithm}

It should be noted that, since each of the updates cannot decrease the value of the objective function, the algorithm is greedy; 
hence, due to the nonlinearity of the objective function, the procedure is guaranteed to converge only to a local optimum.
The iterative routine is stopped when a full iteration yields an increase smaller than a prefixed positive quantity denoted $\texttt{tol}$.

Regarding the hyperparameters, these may be used to introduce prior knowledge in the framework, when this is available.
More in general, as a default, the following values are used: 
the entries of $\bdelta$ are all set to $0.001$, while the entries of $\textbf{a}$ and $\textbf{b}$ are all set to $1$.
With these choices, the $Dirichlet$ distribution on the mixing proportions becomes a shrinkage prior, as already discussed.
The $Gamma(1,1)$ prior on the precisions yields, instead, a rather flexible structure, which supports a wide range of scenarios.
In this paper, the hyperparameters are set to the default values in every application considered.

\subsection{Initialisation}\label{sec:initialisation}
Due to its iterative nature, the greedy algorithm proposed requires the variational parameters to be initialised.
As is the case for other similar algorithms, such as the expectation-maximisation of \textcite{dempster1977maximum} or the iterated conditional modes of \textcite{besag1986statistical},
parameter initialisation plays a fundamental role in avoiding convergence to irrelevant local optima.
Here, a Euclidean-distance-based heuristic method is proposed to initialise the latent positions of the nodes.
Once the value of $K$ is set, the goal is to locate the nodes in $\mathbb{R}^K$ in a way such that the nodes that are close to each other have larger observed interaction values.

First, recall that, given a binary adjacency matrix $\textbf{A}$, 
the generic element indexed by $\left( i,j \right)$ of the matrix $\textbf{AA}'$ identifies the number of paths that allow one to move from $i$ to $j$ in $2$ steps.
Similarly, in a weighted bipartite network context, the matrix $\textbf{XX}'$ has size $M\times M$,
and the value of its generic entry $\left( i,i' \right)$ increases if both senders $i$ and $i'$ are strongly interacting with the same nodes.
Hence, a measure of \textit{similarity} between sender nodes and, respectively, between receiver nodes is introduced using the following matrices:
\begin{equation}
 \textbf{S}_{\textbf{U}} = \sqrt{\frac{\textbf{XX}'}{N}} \hspace{2cm} \textbf{S}_{\textbf{V}} = \sqrt{\frac{\textbf{X}'\textbf{X}}{M}}
\end{equation}
As a consequence, the matrix $\textbf{D}_{\textbf{U}}$ (resp. $\textbf{D}_{\textbf{V}}$), 
obtained inverting the elements of $\textbf{S}_{\textbf{U}}$ (resp. $\textbf{S}_{\textbf{V}}$), 
gauge the dissimilarity between any two sender nodes (resp. receiver nodes).
Gathering together all the available information, a \textit{dissimilarity} matrix for all the nodes of the network is defined as follows:
\begin{equation}
\rowcolors{1}{}{}
 \mathcal{D}=
\left[
\begin{array}{c|c}
\textbf{D}_{\textbf{U}} & \textbf{X} \\
\hline
\textbf{X}' & \textbf{D}_{\textbf{V}}
\end{array}
\right]
\end{equation}
and nonmetric multidimensional scaling is used on $\mathcal{D}$ to initialise the latent positions for all of the senders and receivers.
Since some of the elements of $\textbf{X}$, $\textbf{S}_{\textbf{U}}$ and $\textbf{S}_{\textbf{V}}$ may be zero, 
the initialisation procedure is performed on $\textbf{X}+\textbf{E}$, where the matrix $\textbf{E}$ has all its elements equal to an arbitrary small positive constant $\epsilon>0$.
The variational variances $\tbbeta_{\textbf{U}}$ (resp. $\tbbeta_{\textbf{V}}$) 
are all set to $20$ times the empirical variance of $\tbalpha_{\textbf{U}}$ (resp. $\tbalpha_{\textbf{V}}$).

The remaining variational parameters are all set to $1$ with the exception of the $\tlambda$s, 
which are initialised during the first step of the optimisation conditionally on all the other parameters.

\section{Simulations}\label{sec:simulations}
In this section, the proposed methodology is tested on artificial data, demonstrating that it can serve as a computationally efficient tool to summarise the information provided by nonnegative matrices.

\subsection{Model fit}\label{sec:model_fit}
The goal of these experiments is to demonstrate that the SLPM leads to meaningful results and that it can achieve a good fit to the data.
The NMF method is also used on the same generated datasets for comparison purposes.

\paragraph{Experiment 1: SLPM data.}
In the first experiment, the algorithms are tested on simulated SLPMs in two scenarios: one corresponding to a small network with $M=N=25$, and the other on a larger network with $M=N=100$.
Here, the SLPM modelling assumption is correct; hence, the corresponding method is expected to perform generally well, and to achieve better results than NMF.
The true number of dimensions is fixed to $K_{true}=3$ and the mixing proportions are all set to $1/3$.
The latent positions are all sampled from a standard Gaussian distribution, and, finally, the \textit{average} weighted network $\textbf{X}$ is generated according to:
\begin{equation}\label{eq:gof_1}
x_{ij} = \sum_{k=1}^{K} \lambda_k \left(U_{ik} - V_{jk}\right)^{-2}
\end{equation}
This generating process is independently repeated $100$ times, hence obtaining $100$ simulated SLPMs for each scenario.

The NMF model is estimated using the method of \textcite{brunet2004metagenes}, 
as implemented in the \texttt{R} package \texttt{NMF} \parencite{gaujoux2010flexible}, while the variational method of Section \ref{sec:inference} is used for the SLPM.
Both methods consider $K=8$ latent dimensions, and are run once on each dataset.
Additionally, the SLPM hyperparameters are fixed to $\delta = 0.001$ and $a = b = 1$, 
and the variational parameters are initialised using the distance-based approach described in Section \ref{sec:initialisation}.
The tolerance threshold for the SLPM is set to $\texttt{tol} = 0.01$.

Once the parameters of the SLPM and NMF are estimated for all of the generated networks, 
they are used to reconstruct the weighted adjacency matrices $\textbf{X}^{(SLPM)}$ and $\textbf{X}^{(NMF)}$.
Then, to assess the performance of the two models, the following loss function is considered:
\begin{equation}
 L\left( \textbf{X}, \textbf{X}^{(\cdot)} \right) = \frac{1}{MN}\sum_{i,j} \log\left| x_{ij} - x^{(\cdot)}_{ij} \right|
\end{equation}
for any two $M\times N$ matrices $\textbf{X}$ and $\textbf{X}^{(\cdot)}$.
The presence of the $\log$ is necessary to counterbalance the extremely heavy tailed shape of the weights documented in Figure \ref{fig:sim_dd_12}.
In fact, without the $\log$ transformation, the value of the loss would be almost exclusively determined by the prediction error on the heaviest weight.
By contrast, it should be noted that a drawback of this transformation is that the magnitude of the errors on the small edge values are inflated.

\begin{figure}[htbp]
\centering
\includegraphics[width=0.49\textwidth,page=1]{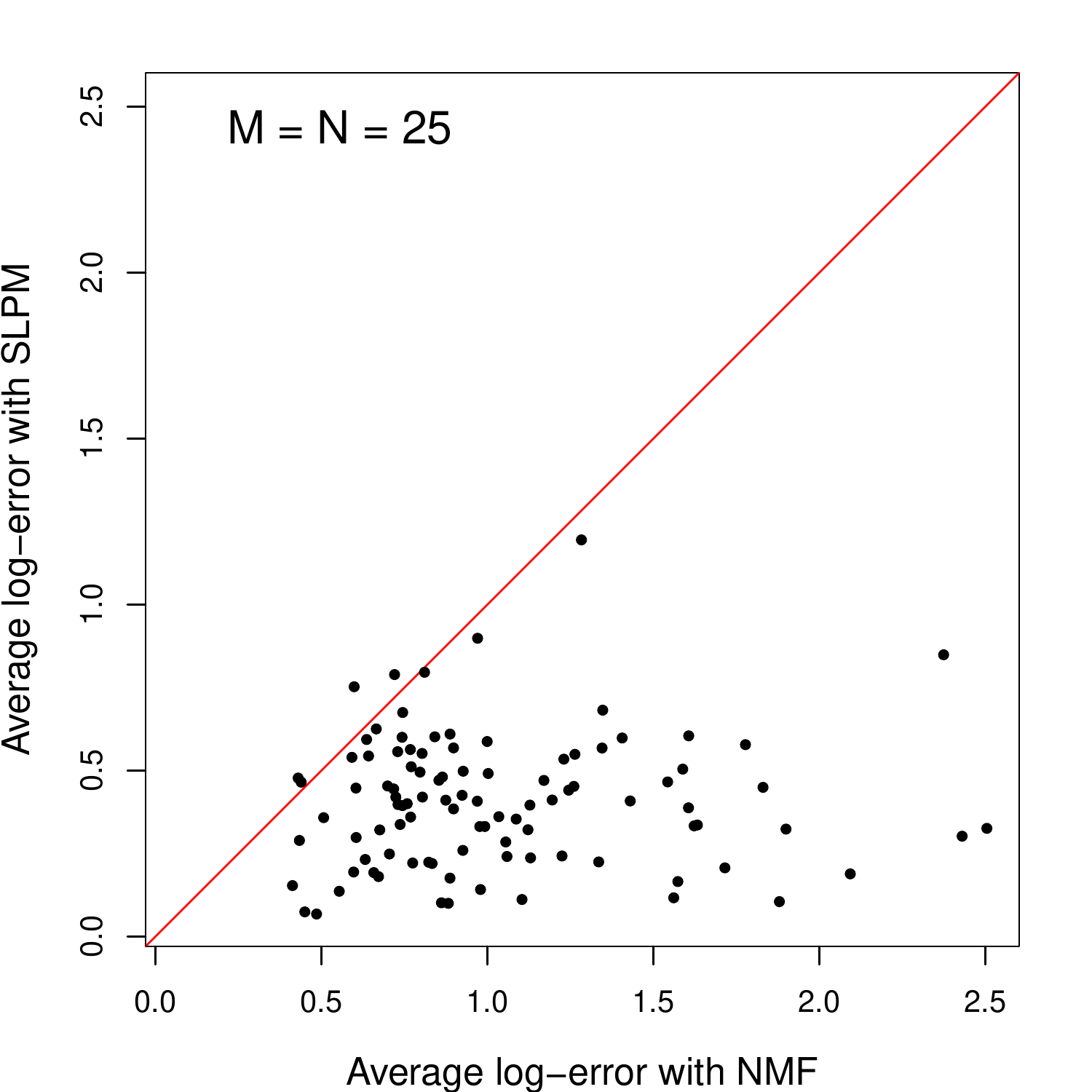}
\includegraphics[width=0.49\textwidth,page=2]{sim_fit_error_1.pdf}
\caption{\textbf{Model fit, experiment 1}. 
Prediction losses of the estimated matrices $\textbf{X}^{(SLPM)}$ and $\textbf{X}^{(NMF)}$ with respect to the true $\textbf{X}$, 
for each of the generated datasets. 
The left panel corresponds to the small datasets ($M=N=25$), whereas the right panel corresponds to the large datasets ($M=N=100$).}
 \label{fig:sim_fit_error_1}
\end{figure}

The two plots in Figure \ref{fig:sim_fit_error_1} compare the losses for SLPM and NMF on the small and large networks, respectively.
Clearly, the estimated SLPM yields smaller prediction $\log$-errors in the vast majority of cases.
This is especially true in the scenario with larger networks (right panel of Figure \ref{fig:sim_fit_error_1}), 
as the additional available information guarantees a reasonably good recovery of the true latent structure.

To estimate the value of $K$, one may consider:
\begin{equation}
 \hat{K} = \#\left\{ k \mbox{ s. t. } k=1,\dots,K; \sum_{i,j} \tlambda_{ijk} > 0\ \right\}
\end{equation}
that is, the number of groups with a size which is strictly greater than zero.
However, due to the presence of extreme edge values, \textit{nearly} empty groups most often appear.
These naturally inflate $\hat{K}$ without adding any substantial changes in the results obtained.
For this reason, studying more directly the mixing proportions may, in general, be advisable.

\begin{figure}[htbp]
\centering
\includegraphics[width=0.49\textwidth,page=1]{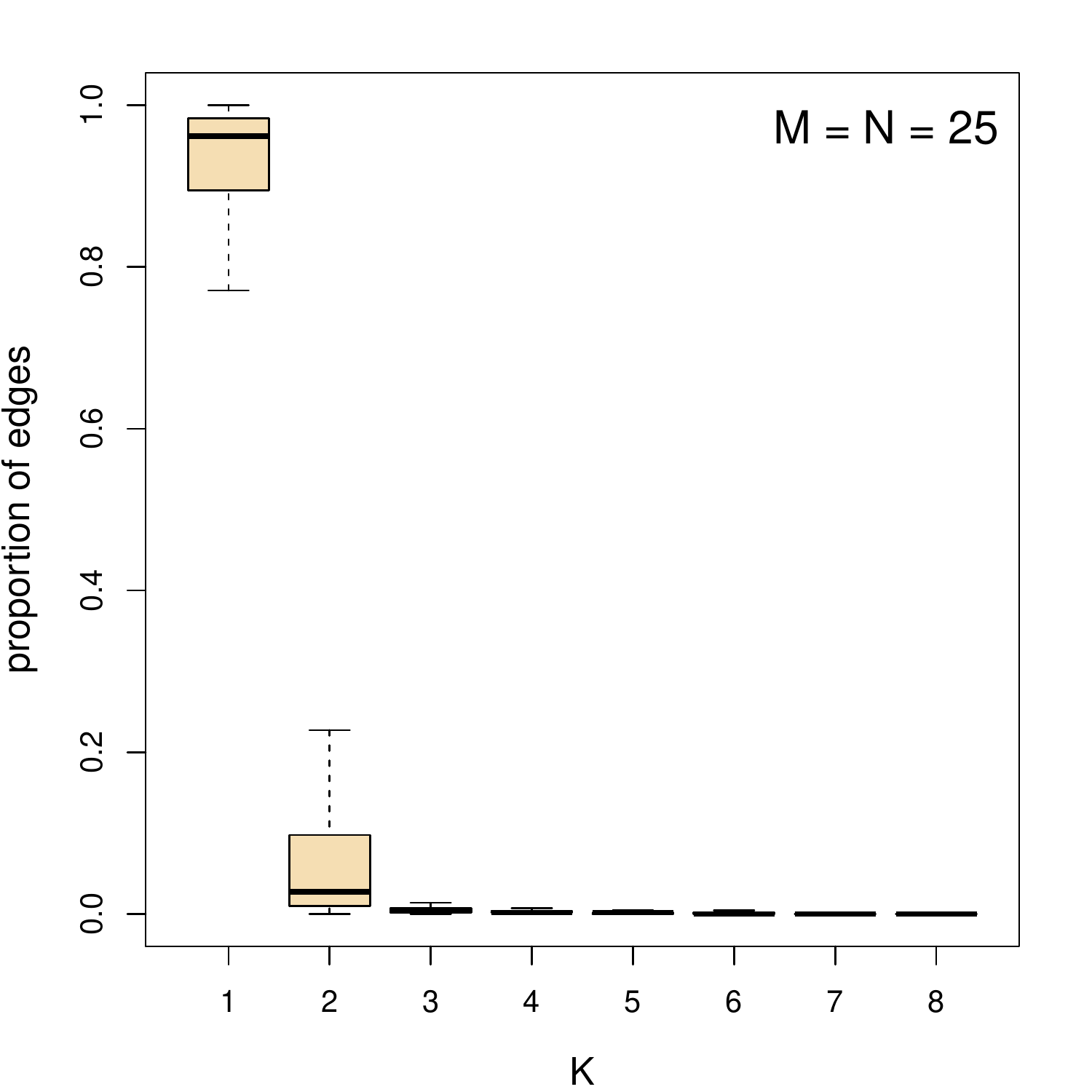}
\includegraphics[width=0.49\textwidth,page=2]{sim_fit_size_1.pdf}
\caption{\textbf{Model fit, experiment 1}. In both panels, the $k$-th boxplot describes the $k$-th mixing weight (in decreasing order) across $100$ datasets.
In both small networks (left panel) and large networks (right panel), two latent dimensions stand out.}
 \label{fig:sim_fit_size_1}
\end{figure}

Regarding the estimation of $K$ in this experiment, Figure \ref{fig:sim_fit_size_1} shows the estimated mixing proportions in decreasing order.
In both scenarios, and for the vast majority of generated datasets, two latent dimensions stand out, since most edges are assigned to either of these two.
The third most relevant dimension generally contains only few edges which correspond to outliers.

\begin{figure}[htbp]
\centering
\includegraphics[width=0.49\textwidth]{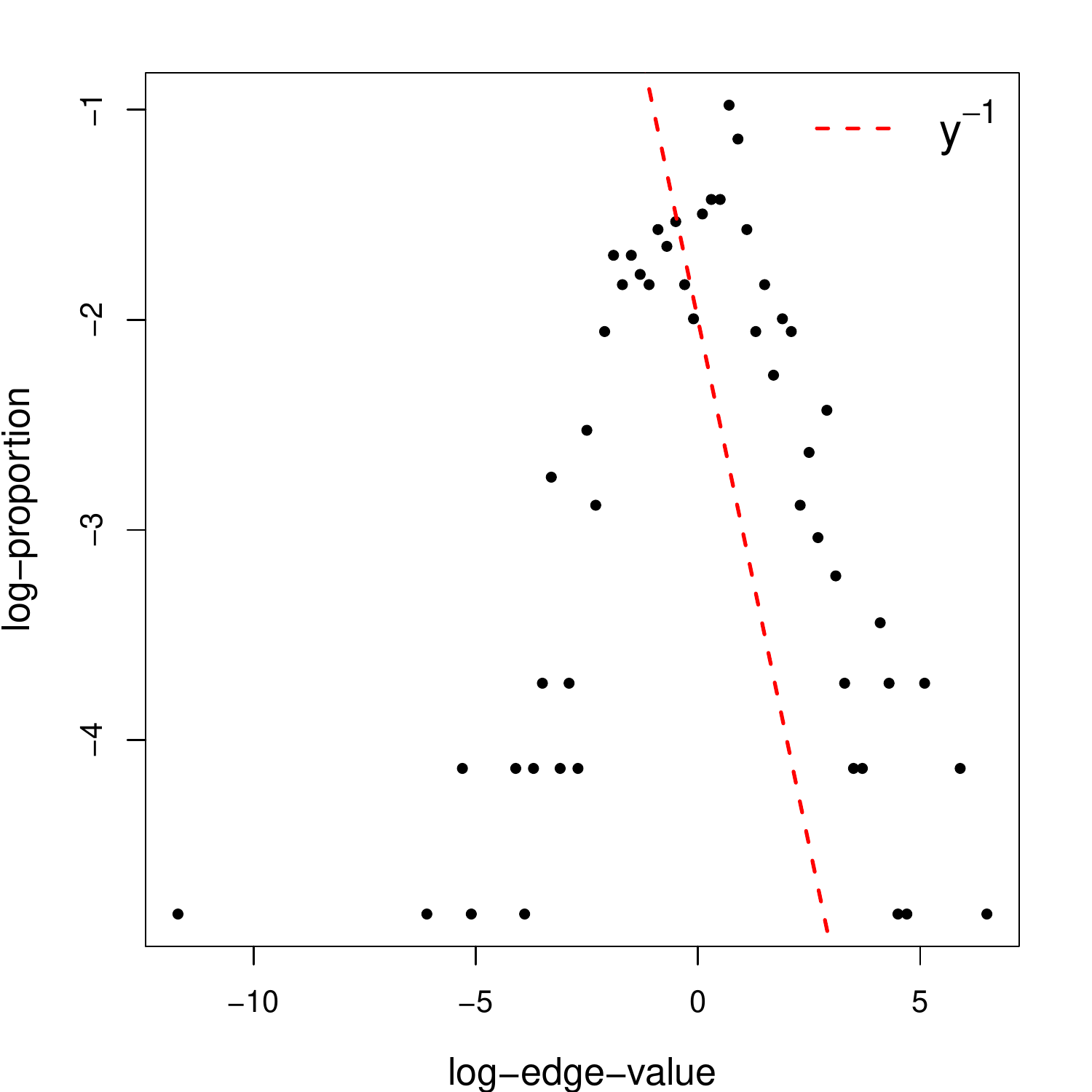}
\includegraphics[width=0.49\textwidth]{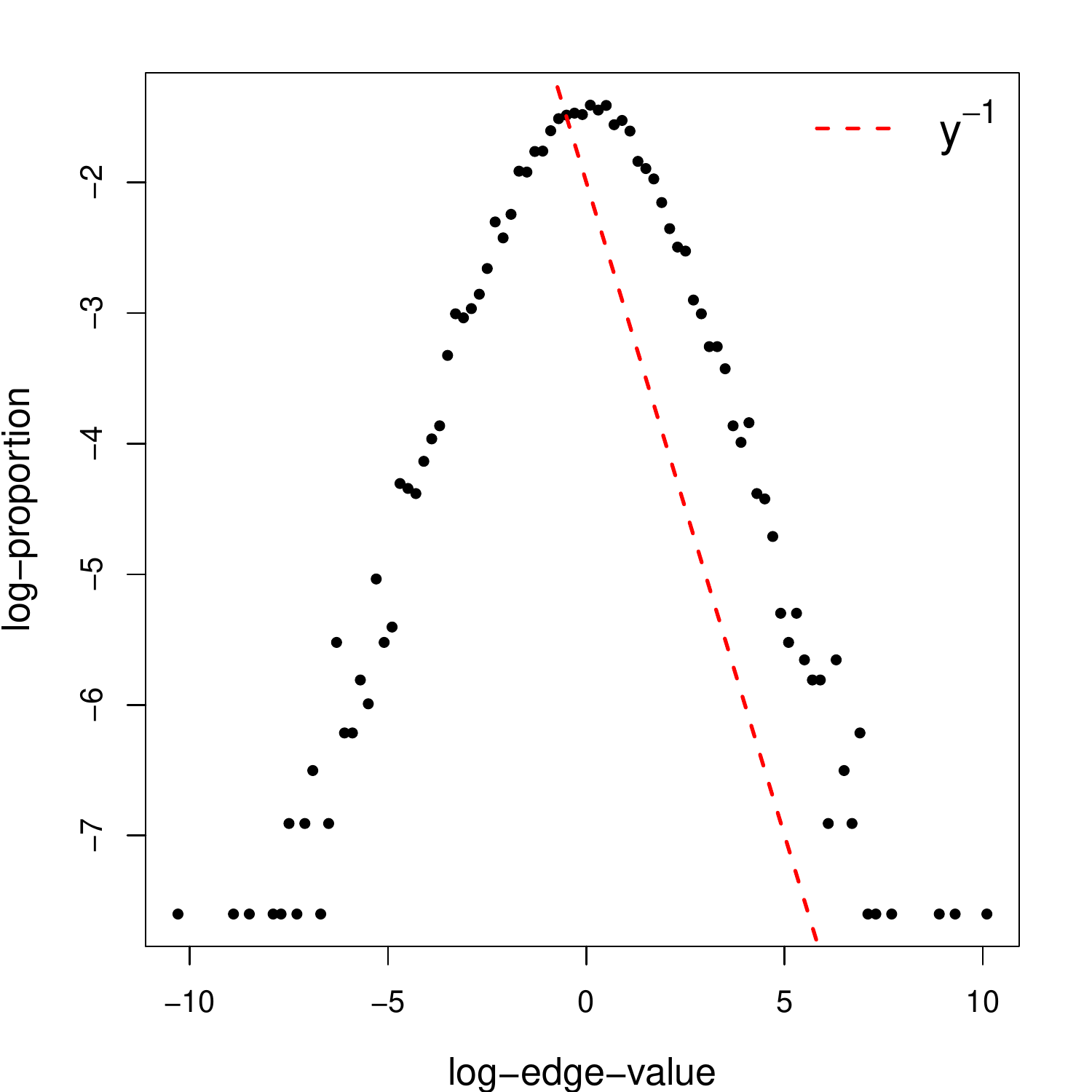}
\caption{\textbf{Model fit, experiment 2}. 
Empirical $\log$-$\log$ distribution of the edge weights for two networks generated using the homogeneous model, with sizes $M=N=25$ and $M=N=100$ (respectively on the left and right panel).}
 \label{fig:sim_dd_34}
\end{figure}

\paragraph{Experiment 2: homogeneous data.}
In this second experiment, the data is generated from a homogeneous process: 
each edge weight is assumed to be an exponential random variable, whose rate is distributed according to a $Gamma(1,1)$.
This leads to a homogeneous network model whose empirical degree distribution is characterised by a heavy tail (Figure \ref{fig:sim_dd_34}).

Similarly to the previous experiment, the SLPM and NMF are estimated on $100$ dataset with sizes $M=N=25$ or $M=N=100$.
The hyperparameters, number of latent dimensions, and optimisation parameters are also set exactly to the same values used in the previous experiment.

\begin{figure}[htbp]
\centering
\includegraphics[width=0.49\textwidth,page=1]{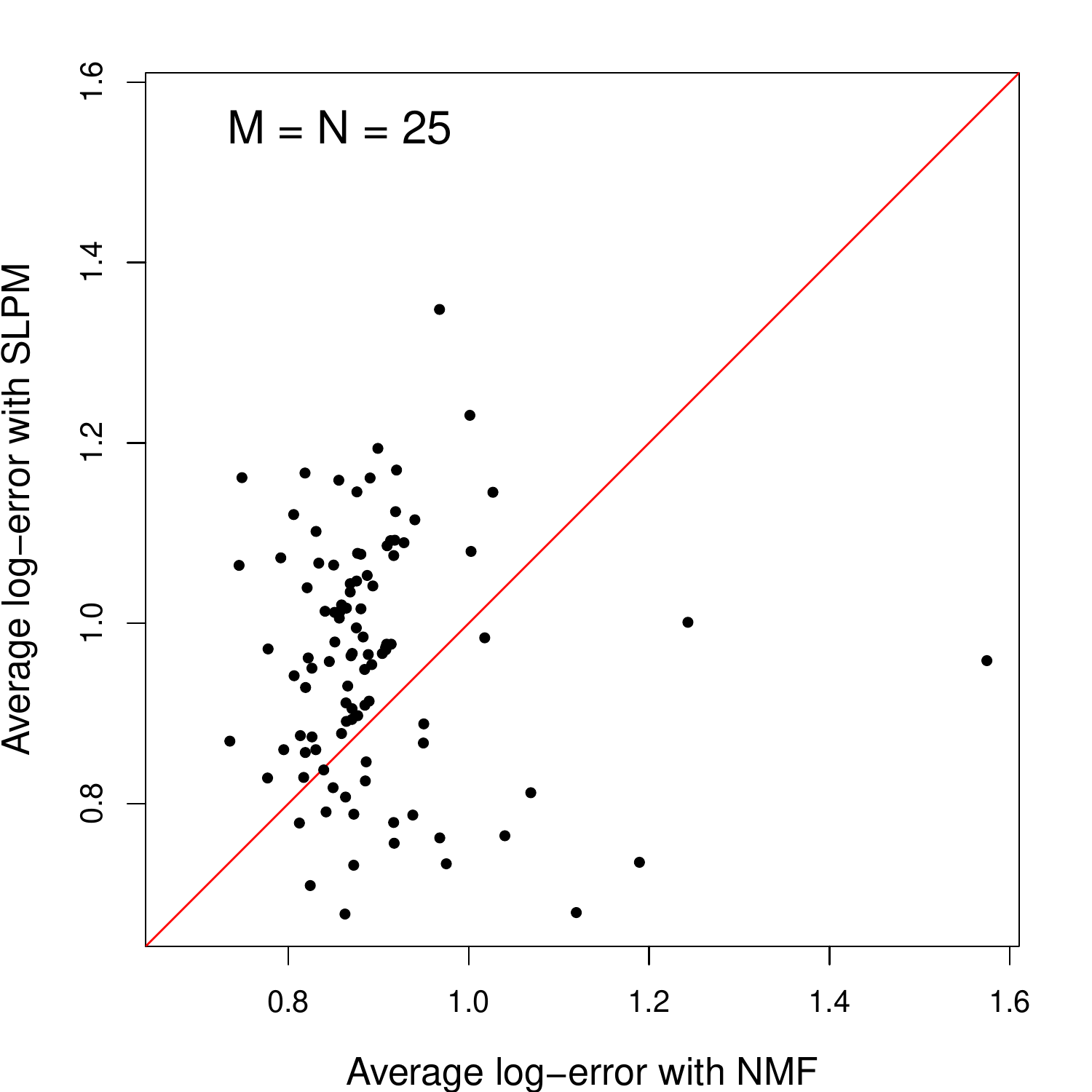}
\includegraphics[width=0.49\textwidth,page=2]{sim_fit_error_2.pdf}
\caption{\textbf{Model fit, experiment 2}. 
Prediction losses of the estimated matrices $\textbf{X}^{(SLPM)}$ and $\textbf{X}^{(NMF)}$ with respect to the true $\textbf{X}$, 
for each of the generated datasets. 
The left panel corresponds to the small datasets ($M=N=25$), whereas the right panel corresponds to the large datasets ($M=N=100$).}
 \label{fig:sim_fit_2}
\end{figure}

The predictive $\log$-errors are shown in Figure \ref{fig:sim_fit_2}.
While the NMF model seems to achieve slightly better results on the small datasets, the SLPM again outperforms NMF on the larger datasets.
This suggests that the distance-based generative process behind the SLPM may be more flexible than that of the NMF, 
and that it may be better suited to represent the topologies exhibited by many types of observed networks.

\begin{figure}[htbp]
\centering
\includegraphics[width=0.49\textwidth,page=1]{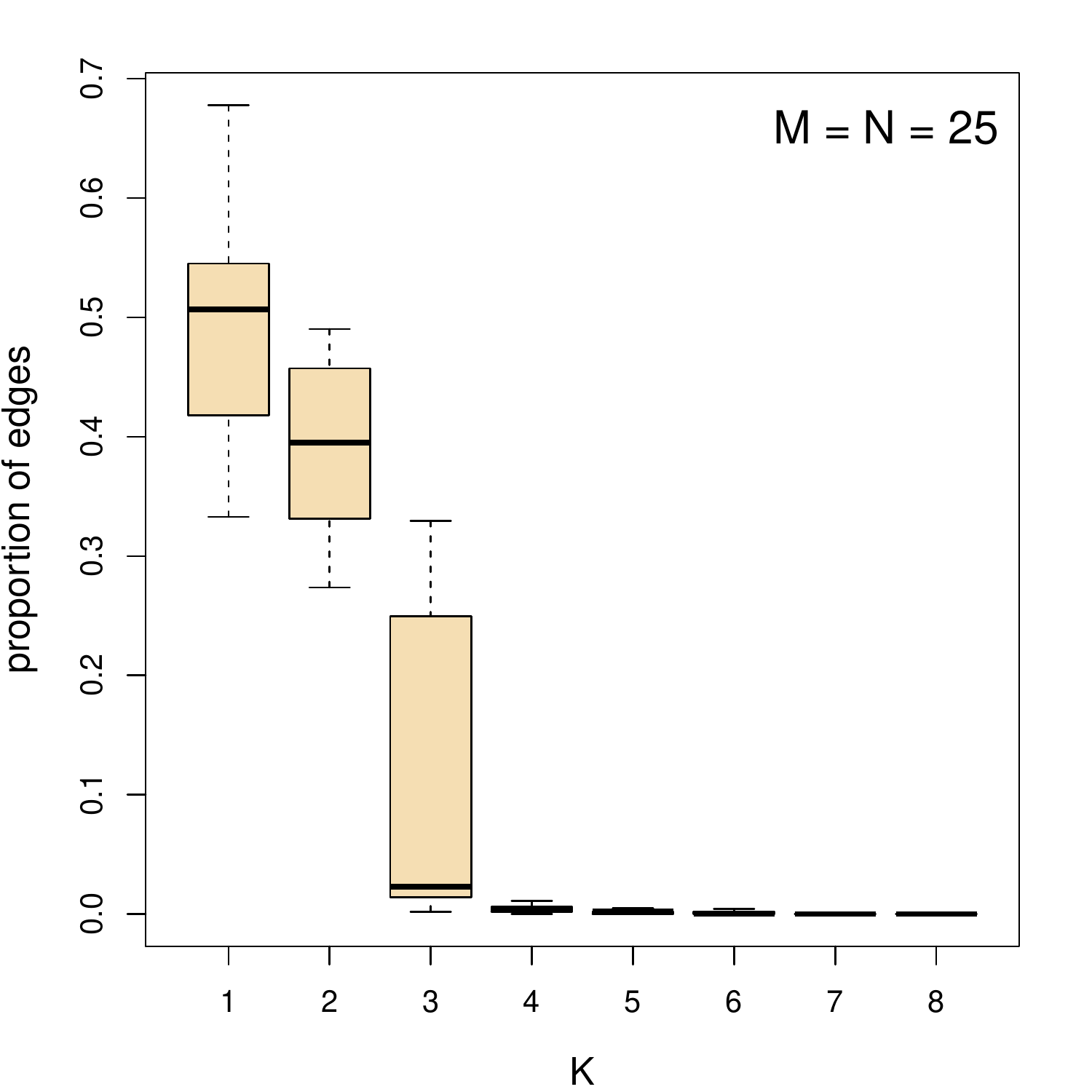}
\includegraphics[width=0.49\textwidth,page=2]{sim_fit_size_2.pdf}
\caption{\textbf{Model fit, experiment 2}. In both panels, the $k$-th boxplot describes the $k$-th mixing weight (in decreasing order) across $100$ datasets.
In the small networks (left panel) the estimated models are characterised by three latent dimensions, 
whereas in the large networks (right panel) more dimensions are required to fit the data.}
 \label{fig:sim_fit_size_2}
\end{figure}

As concerns the estimation of $K$, Figure \ref{fig:sim_fit_size_2} shows the estimated mixing proportions in decreasing order.
The number of relevant components corresponds to the correct number of dimensions in the small datasets, 
since we see from the left panel that the third largest group contains approximately five to $30\%$ of the edges in more than half of the datasets.
However, an overestimation of $K$ is observed on the larger dataset.
Here, it seems that, for most datasets, the edges are allocated in approximately similar numbers to any of five or six groups.
This type of behaviour may be justified by the model misspecification, 
in that the latent positions are not as effective in explaining the variance in the data, 
and more latent dimensions become necessary to achieve a good fit.

\subsection{Model choice}\label{sec:model_choice}
The goal of this section is to further assess the performance of the SLPM in recovering the correct number of latent dimensions.

Four scenarios are considered: 
$K_{true}=2$ and $M=N=40$ in the first scenario, 
$K_{true}=4$ and $M=N=40$ in the second, 
$K_{true}=2$ and $M=N=80$ in the third, 
and $K_{true}=4$ and $M=N=80$ in the fourth. 
In each scenario the hyperparameters are all fixed to $1$, and $100$ artificial networks are generated independently.
The mixing proportions, the precisions, and the latent positions are all generated according to the hierarchical structure described in Section \ref{sec:SLPM_model}.
Then, the \textit{average} weighted network $\textbf{X}$ is generated according to \eqref{eq:gof_1}.

Similarly to the previous section, 
the SLPM is estimated on all of the datasets using VB with the same set-up for hyperparameters, number of latent dimensions, and optimisation parameters.
In addition, in order to assess the sensitivity of the SLPM to its hyperparameters, 
the variational algorithm is run once with $\delta = 0.1$ and once with $\delta = 0.001$.

\begin{figure}[htbp]
\centering
\includegraphics[width=0.49\textwidth,page=1]{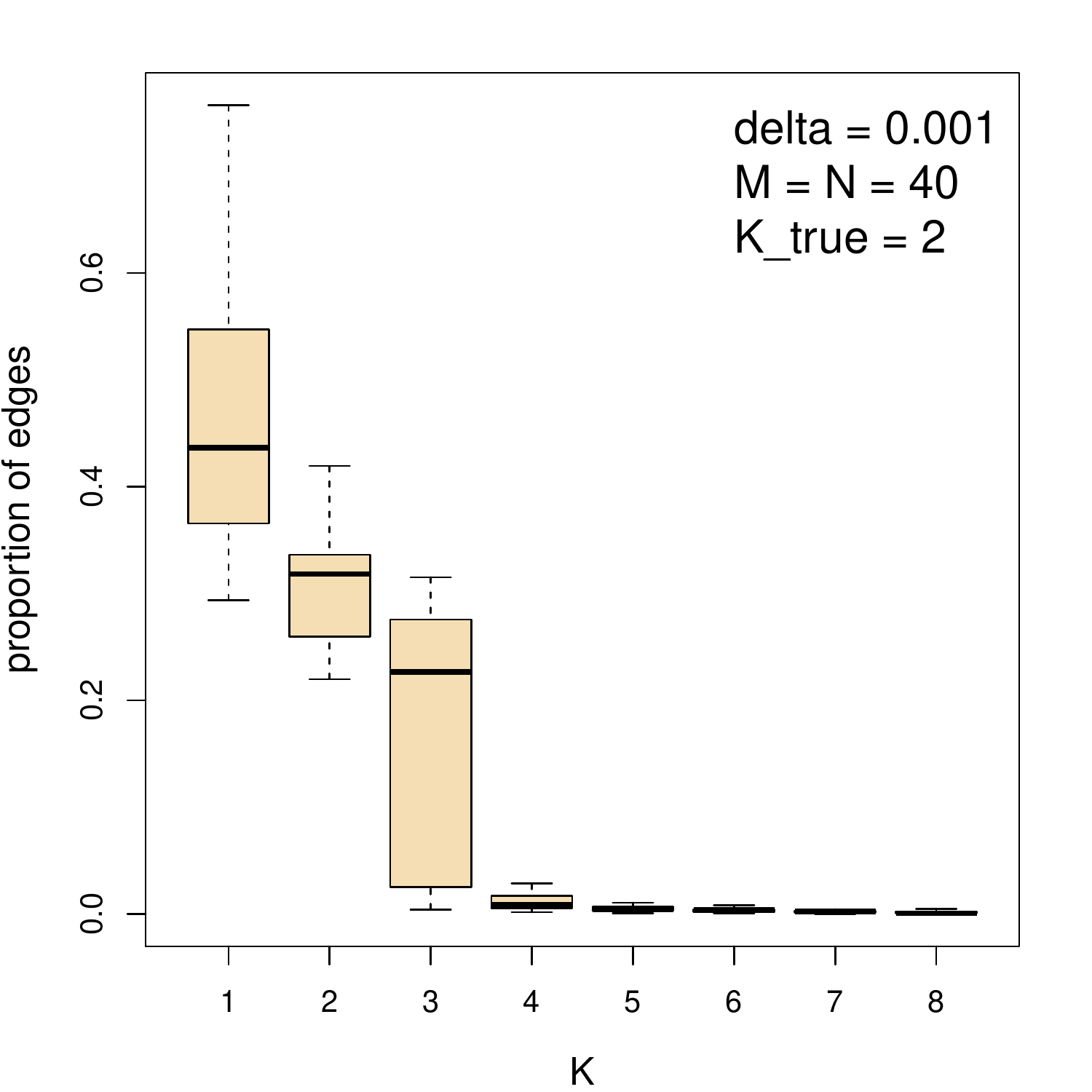}
\includegraphics[width=0.49\textwidth,page=2]{sim_choice_1.pdf}
\caption{\textbf{Model choice, scenario 1}. In both panels, the $k$-th boxplot describes the $k$-th mixing weight (in decreasing order) across $100$ datasets.
The hyperparameter $\delta$ seems to have no effect on the results. Three latent dimensions stand out with a weight approximately equal to $0.2$ in at least half of the datasets. 
The fourth dimension has a small weight in all of the datasets.}
 \label{fig:sim_choice_1a}
\end{figure}

\begin{figure}[htbp]
\centering
\includegraphics[width=0.49\textwidth,page=3]{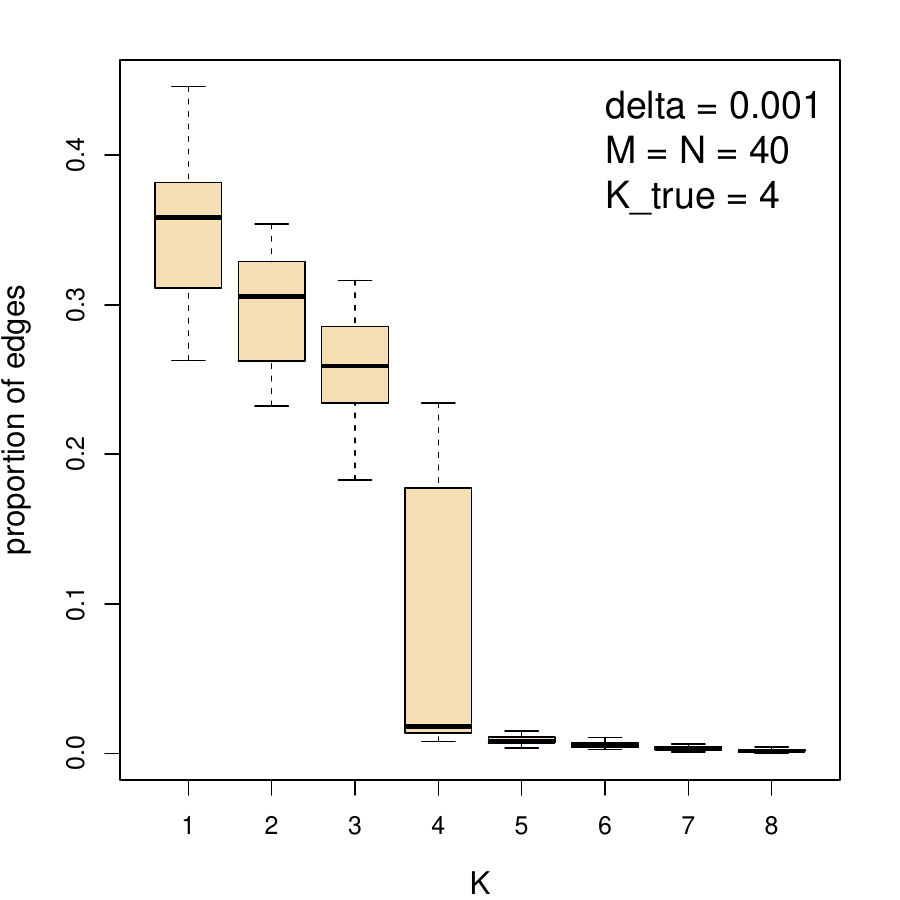}
\includegraphics[width=0.49\textwidth,page=4]{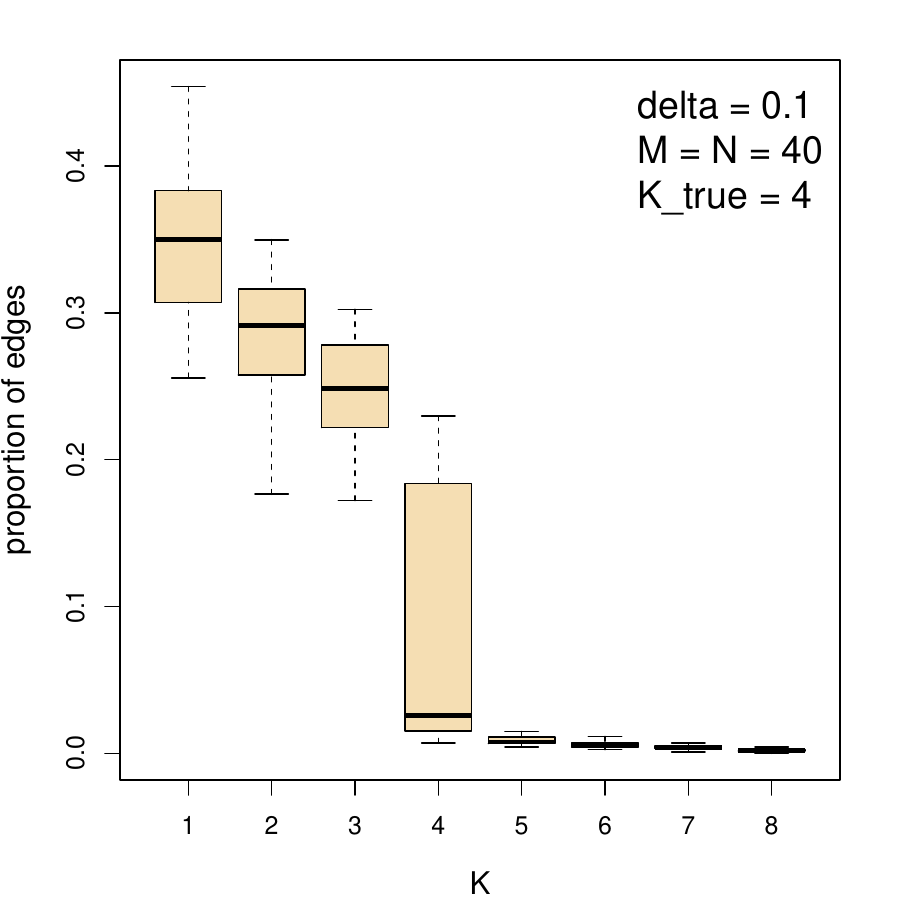}
\caption{\textbf{Model choice, scenario 2}. In both panels, the $k$-th boxplot describes the $k$-th mixing weight (in decreasing order) across $100$ datasets.
The hyperparameter $\delta$ seems to have no effect on the results. In this scenario, the number of relevant latent dimensions is generally equal to the true value of $K$.}
 \label{fig:sim_choice_1b}
\end{figure}

\begin{figure}[htbp]
\centering
\includegraphics[width=0.49\textwidth,page=5]{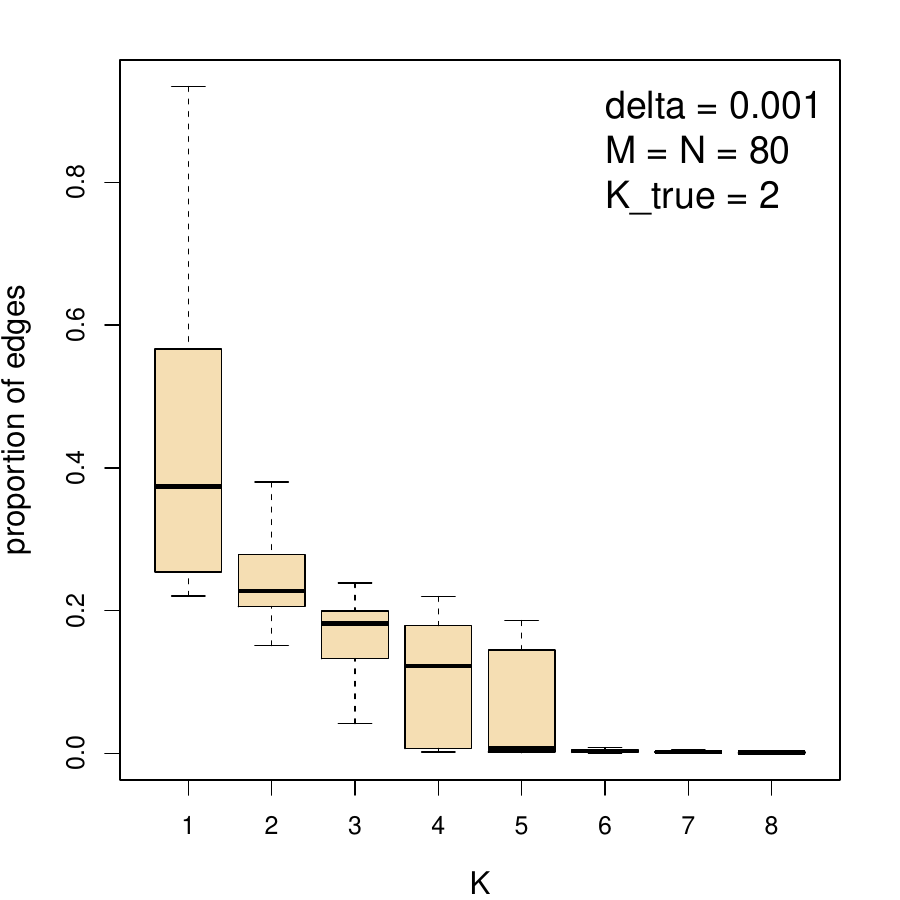}
\includegraphics[width=0.49\textwidth,page=6]{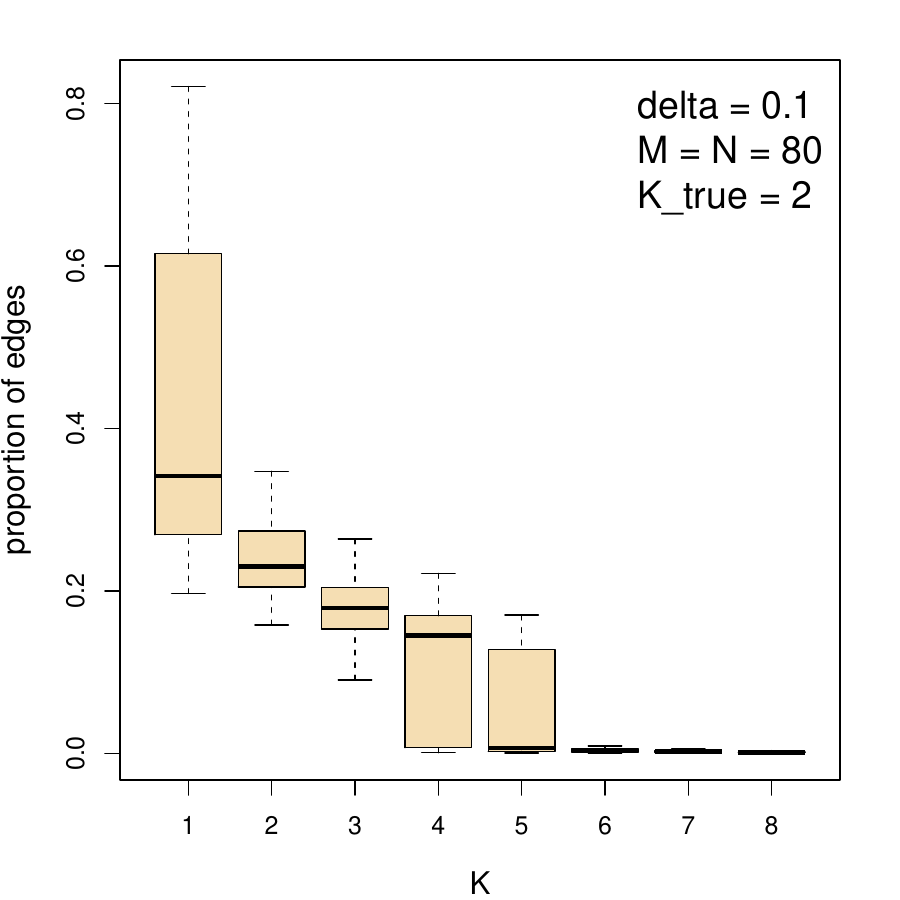}
\caption{\textbf{Model choice, scenario 3}. In both panels, the $k$-th boxplot describes the $k$-th mixing weight (in decreasing order) across $100$ datasets.
The hyperparameter $\delta$ seems to have no effect on the results. Although the true value of $K$ is two, five latent dimensions stand out. 
In this scenario, the VB algorithm struggles to reduce the dimensionality and to recover the correct number of latent dimensions.}
 \label{fig:sim_choice_1c}
\end{figure}

\begin{figure}[htbp]
\centering
\includegraphics[width=0.49\textwidth,page=7]{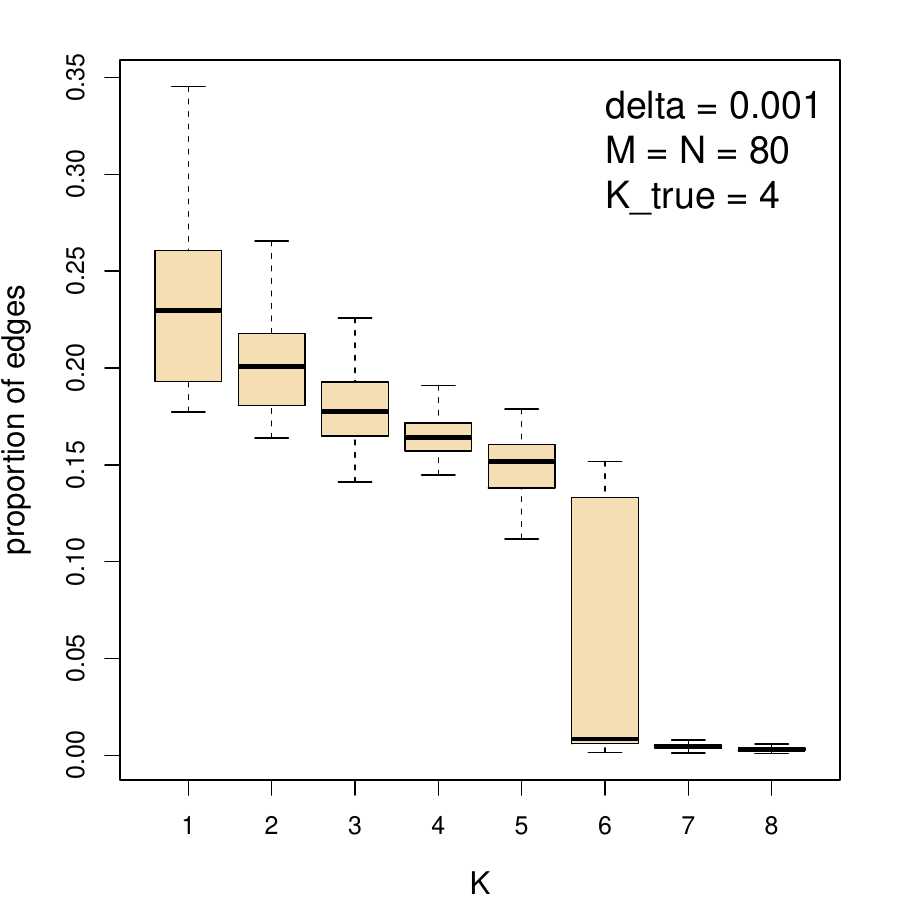}
\includegraphics[width=0.49\textwidth,page=8]{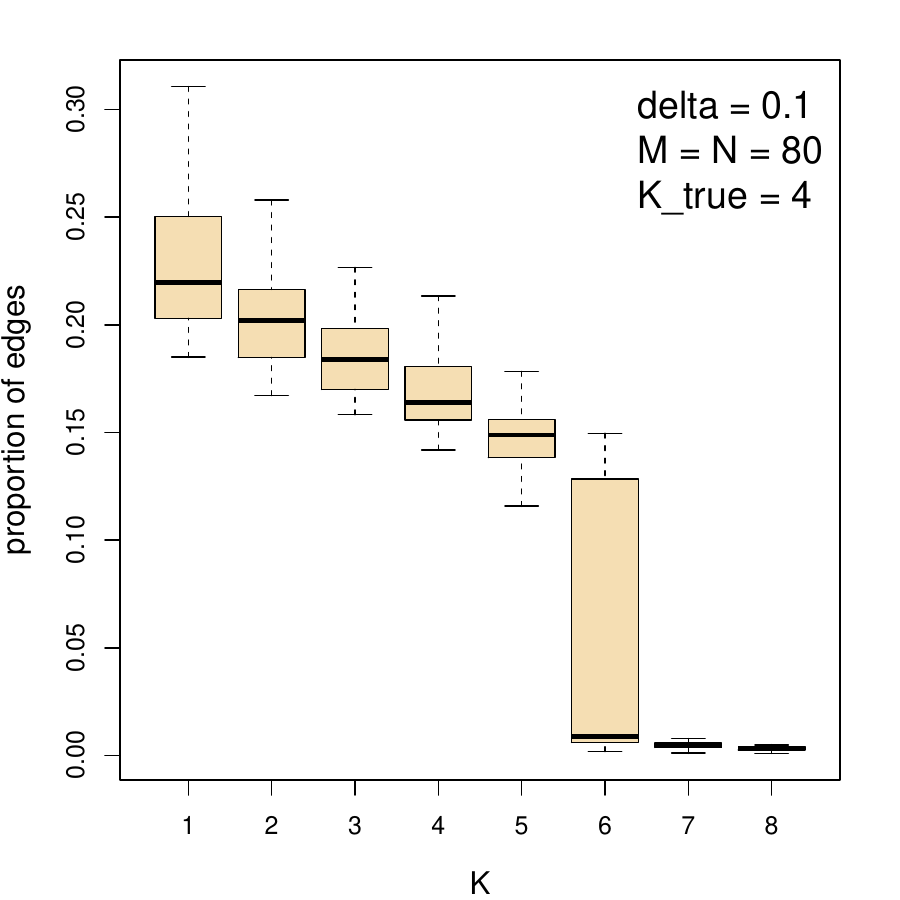}
\caption{\textbf{Model choice, scenario 4}. In both panels, the $k$-th boxplot describes the $k$-th mixing weight (in decreasing order) across $100$ datasets.
The hyperparameter $\delta$ seems to have no effect on the results. 
Similarly to the third scenario, the VB algorithm struggles to reduce the dimensionality and it tends to overestimate the number of relevant latent dimensions.}
 \label{fig:sim_choice_1d}
\end{figure}

As in the previous simulation studies, the number of revelant groups is deduced by the mixing proportions once they are sorted in decreasing order.
Figures \ref{fig:sim_choice_1a}, \ref{fig:sim_choice_1b}, \ref{fig:sim_choice_1c}, and \ref{fig:sim_choice_1d} refer to scenarios one to four, respectively.
It appears that, overall, the value of the hyperparameter $\delta$ does not have any relevant effect on the estimate of $K$.
While in the first scenario ($M=N=40$, $K=2$, Figure \ref{fig:sim_choice_1a}) the number of relevant latent dimensions is generally overestimated to $3$,
in the second scenario ($M=N=40$, $K=4$, Figure \ref{fig:sim_choice_1b}) $K$ is well-recovered.
In fact, in more than half of the datasets, the fourth group contains more than $5\%$ of the edges, whereas the fifth group has negligible size.

Similarly to the previous experiment, $K$ is overestimated in the larger datasets 
($M=N=80$, scenarios $3$ and $4$ in Figures \ref{fig:sim_choice_1c} and \ref{fig:sim_choice_1d}, respectively),
in that in more than half of the datasets the sixth group has a non-negligible size.

These simulations illustrate that, while shrinkage and parsimony are generally achieved, 
the estimated number of relevant groups does not always coincide exactly with the true $K$, but rather lives in a neighourhood of this value.
In particular, the number of relevant groups tends to be overestimated when the dataset is large and, hence, more heterogeneous.

\subsection{Computational efficiency}\label{sec:computational efficiency}
Similarly to the other available estimation methods for LPMs, the computational complexity of the variational algorithm is determined by the updates of the latent positions.
One such update requires a summation over all possible neighbours, hence it implies either $N$ or $M$ operations, depending on whether the updated node is a sender or receiver, respectively.
In addition, since the step size must be tuned at every update attempt, 
the $N$ (or $M$) evaluations are actually performed an unspecified number of times, depending on the rule determining the learning rates.
However, since the step sizes remain approximately constant as the algorithm progresses, these re-evaluations are generally performed very few times.

The number of iterations required by the algorithm to converge impacts the overall computing time by a much greater extent.
In general, although the algorithm is greedy, there are no guarantees regarding the number of iterations required for convergence, and, 
most likely, such number depends on the number of nodes and on the topology of the network.
In practice, a few hundreds iterations are often sufficient to achieve the threshold $\texttt{tol} = 0.01$.

\begin{figure}[htbp]
\centering
\includegraphics[width=0.49\textwidth]{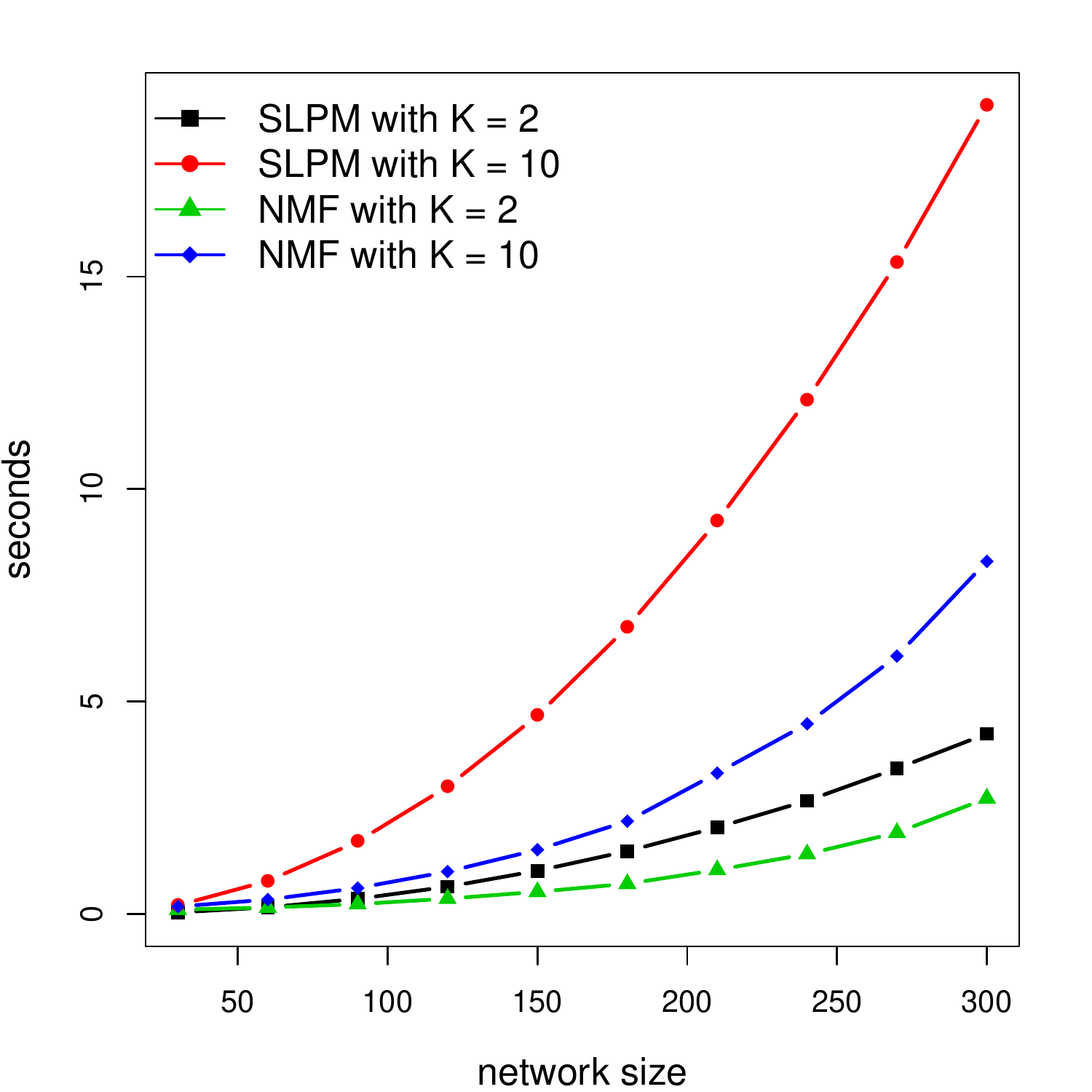}
\caption{\textbf{Computational efficiency}. 
Number of seconds required to perform the first $10$ iterations of the VB algorithm (black and red lines), and one run of the NMF algorithm (green and blue lines).
The two algorithms appear to scale similarly with respect to the size of the networks, especially when the number of latent dimensions is small. 
For a larger number of latent dimensions, the NMF algorithm scales much better than the VB algorithm used to estimate the SLPM.}
 \label{fig:sim_time_1}
\end{figure}

Figure \ref{fig:sim_time_1} compares the computing time required by the variational algorithm of SLPM with that of NMF.
NMF exploits parallel computing and is generally much faster, as it converges with very few iterations.
The figure also highlights that SLPM scales quite poorly with the number of dimensions.

These results suggest that, while SLPM generally achieves a very good fit to the data, 
it may also struggle in dealing with very large datasets, hence forcing one to increase the tolerance threshold or to reduce the maximum number of iterations,
in order to get usable results in a reasonable time.

\section{Real data applications}\label{sec:applications}
The goal of the two studies shown in this section is to propose an application of the methodology proposed on real datasets, 
and to illustrate how the SLPM represents and summarises the observed data.

\subsection{Hospital encounter data}
The dataset used in this example has been first studied by \textcite{vanhems2013estimating}, 
and it is available from the \texttt{R} package \texttt{igraphdata}.
The data record proximity interactions between $29$ patients, $27$ nurses, $11$ doctors and $8$ administrative staff at a university Hospital in Lyon, France, 
from a Monday to the following Friday in December $2010$.
Wearable sensors based on active Radio-Frequency IDentification (RFID) are used to probe the area surrounding each individual at $20$ seconds intervals,
and proximity interactions are recorded whenever two individuals are closer than $1.5$ meters.
The analysis of such data may provide insights on the dynamics of the spread of Hospital-acquired infections, 
and, hence, lead to valid and more effective preventive measures.

In this analysis, the interaction matrix $\textbf{X}$ denotes the cumulated proximity time, 
i.e. the generic entry $x_{ij}$ indicates the number of seconds the staff member $i$ has spent in proximity of patient $j$.
The length of stay of the patients can be deduced from the proximity information, and, clearly, those who stayed longer have a higher degree (not shown).

The latent number of dimensions was set to $K=10$, and the VB algorithm was run once using the initialisation of Section \ref{sec:initialisation}.
The algorithm reached convergence after $245$ iterations and $5.7$ seconds.

\begin{figure}[htbp]
 \centering
 \includegraphics[width=0.49\textwidth]{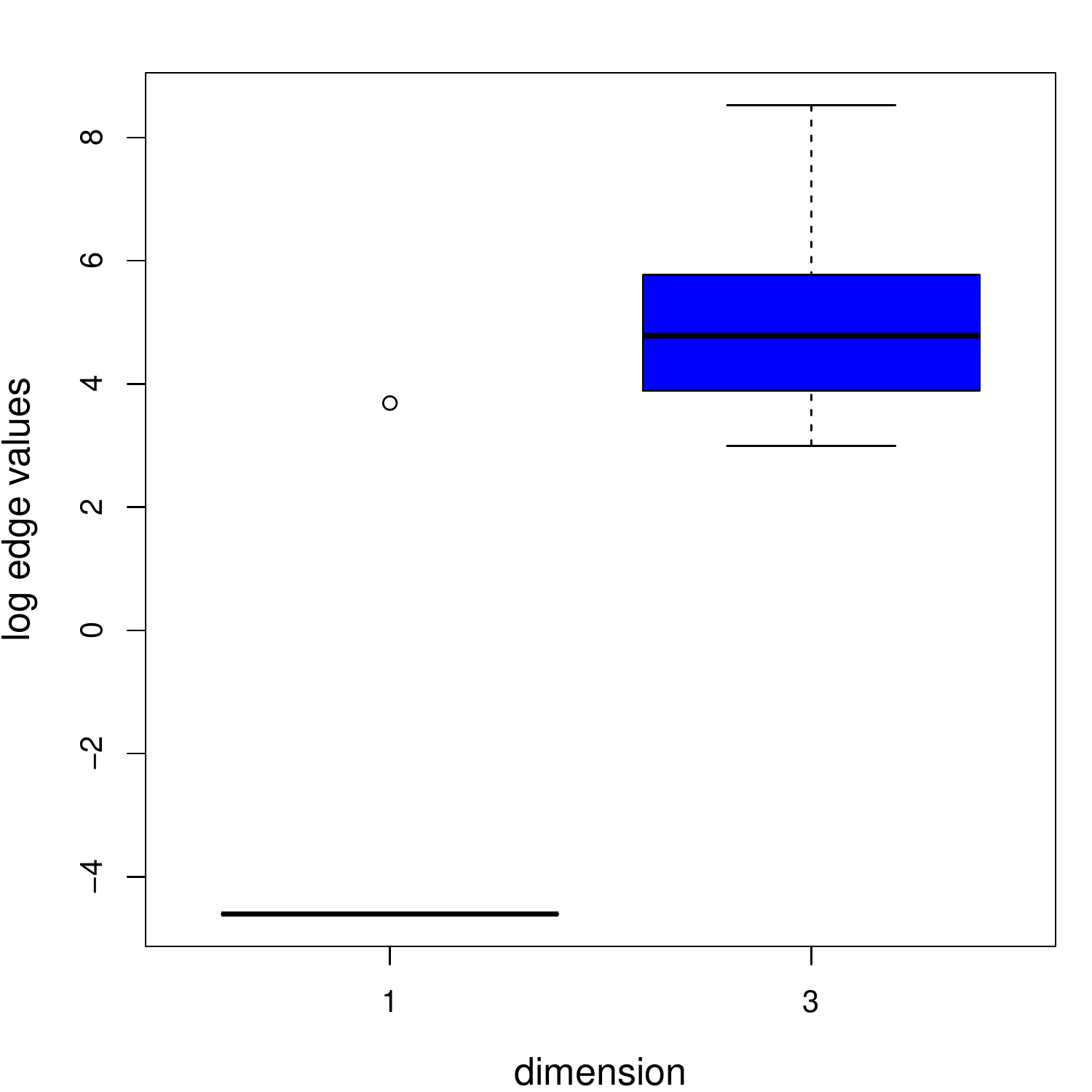}
 \includegraphics[width=0.49\textwidth]{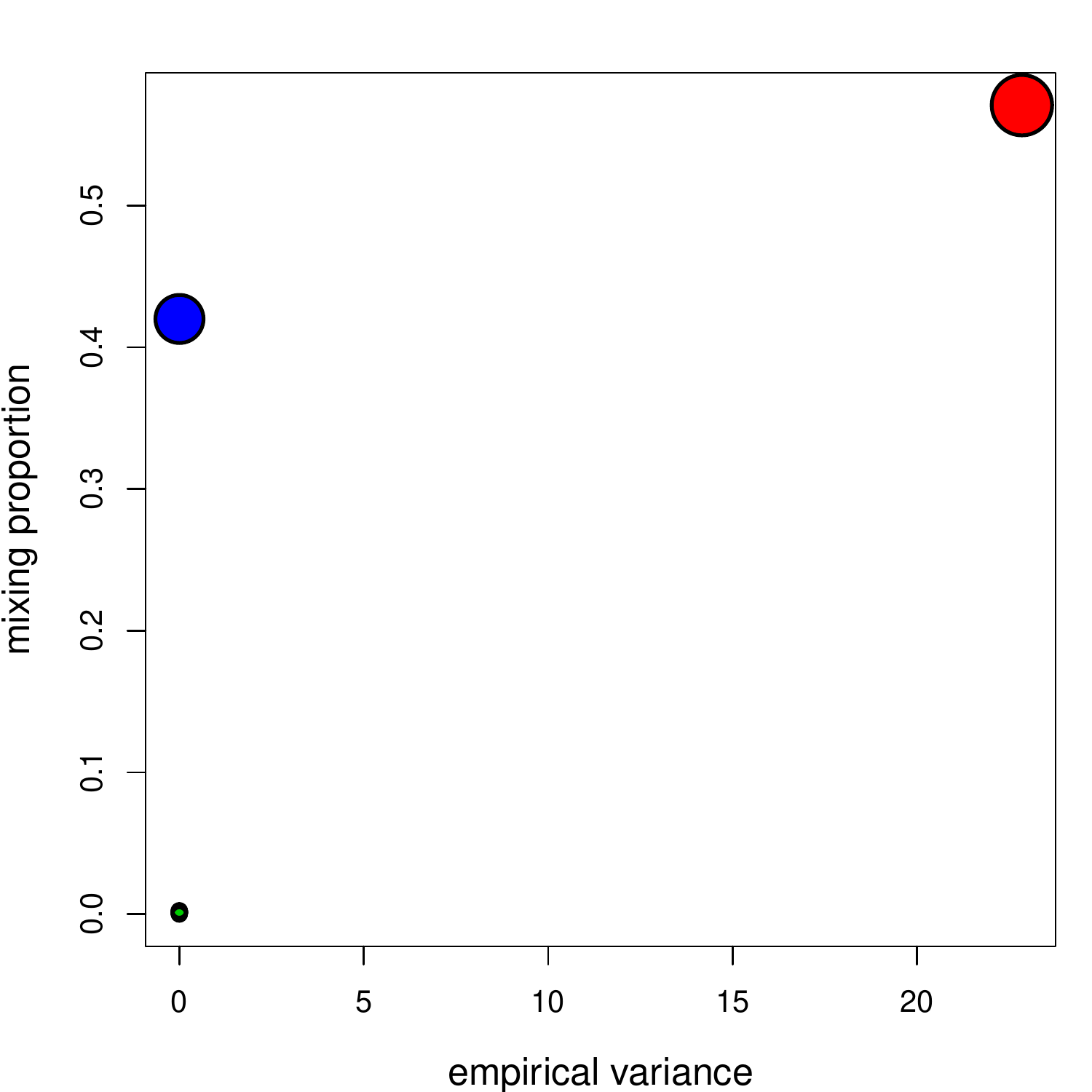}
 \caption{\textbf{Hospital data}. The left panel shows the distribution of the weight log-values assigned to each of the two most relevant dimensions.
 As null-weights cannot be shown, a small positive constant is added to all the entries of $\textbf{X}$ to improve the readability of this plot.
 The right panel compares the mixing proportions with the empirical variances of the corresponding latent dimensions.}
 \label{fig:rfid_a}
\end{figure}

\begin{figure}[ht]
 \centering
 \includegraphics[width=0.6\textwidth]{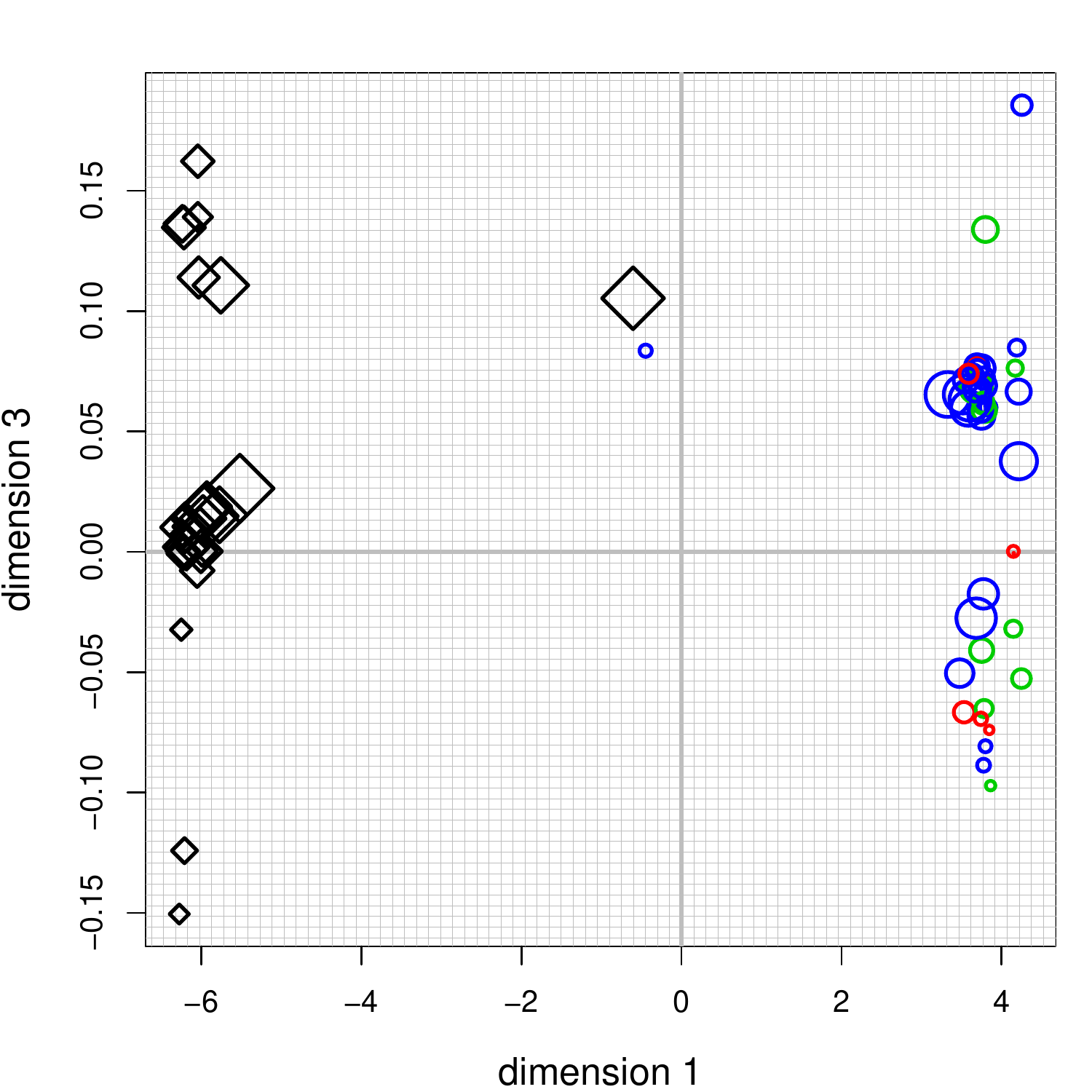}
 \caption{\textbf{Hospital data}. Latent space representation based on the two most relevant latent dimensions.
 The black squares correspond to the patients, whereas the circles correspond to the staff members.
 Different circle colours indicate the $3$ known staff member occupations (blue for nurses, green for doctors, red for administrative staff).
 The sizes of the circles and squares are proportional to the degrees of the corresponding nodes.}
 \label{fig:rfid_b}
\end{figure}

The rounded mixing proportions are $\hat{\lambda}_1 = 0.571$, $\hat{\lambda}_3 = 0.420$, and $\hat{\lambda}_k < 0.003$ for $i=2,5,6,7,8,9,10$;
two latent groups stand out with a large mixing weight.
As illustrated by the two plots in Figure \ref{fig:rfid_a}, the two relevant latent dimensions fulfil two very different roles.
In fact, the larger one has a very high empirical variance, meaning that it is used to represent all of the zero or nearly-zero weights.
By contrast, dimension number three is instead used for almost all of the remaining edges.
The same conclusion is also apparent from Figure \ref{fig:rfid_b}, where dimension number three, plotted on the vertical axis, 
clearly exhibits more heterogeneity than dimension number one, which is plotted on the horizontal axis.

The results show that there are no substantial differences between the nurses, doctors, and administrative staff in how they interact with the patients.
It should be noted that, since dimension three is able to capture almost all the heterogeneity in the data, 
the unidimensional projection of this SLPM has features very similar to those of the original LPM of \textcite{hoff2002latent}.
For example, for both patients and staff, nodes with higher degree tend to be positioned closer to the centre of the space.

\subsection{Microarray data}
The second example is an analysis of the dataset first studied by \textcite{khan2001classification}.
Here, the matrix $\textbf{X}$, of size $2308\times 64$, represents the measured expression level of $M=2308$ genes in $N=64$ samples, 
for patients who tested positive on the presence of small round blue cell tumours of childhood.
The goal is to characterise the connectivity profiles of genes and samples, and to provide a latent space representation of these data.
There are no zero-weighted edges, the lowest value recorded is $0.0025$, the highest is $32.66$, and the median is $0.5666$.

\begin{figure}[htbp]
 \centering
 \includegraphics[width=0.49\textwidth]{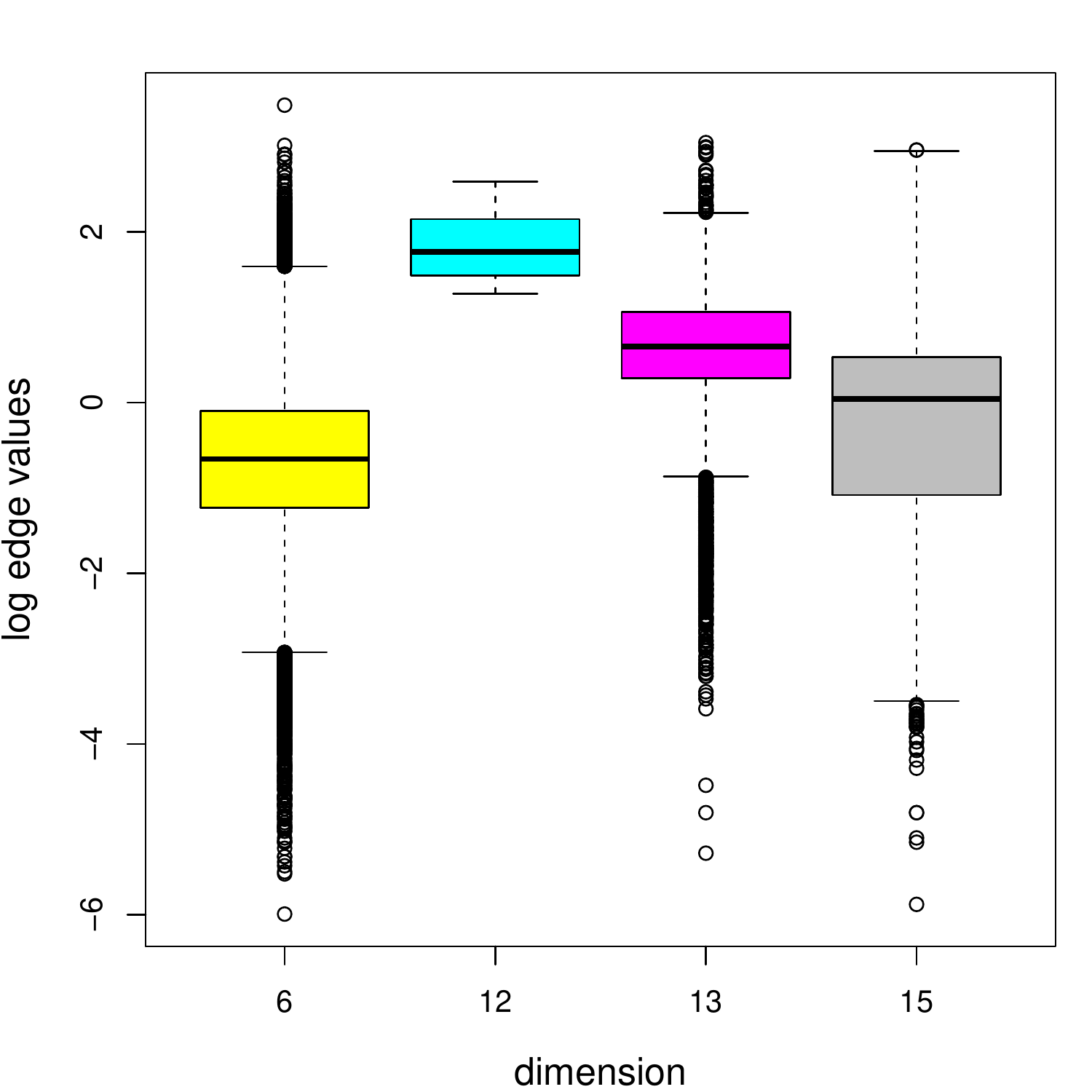}
 \includegraphics[width=0.49\textwidth]{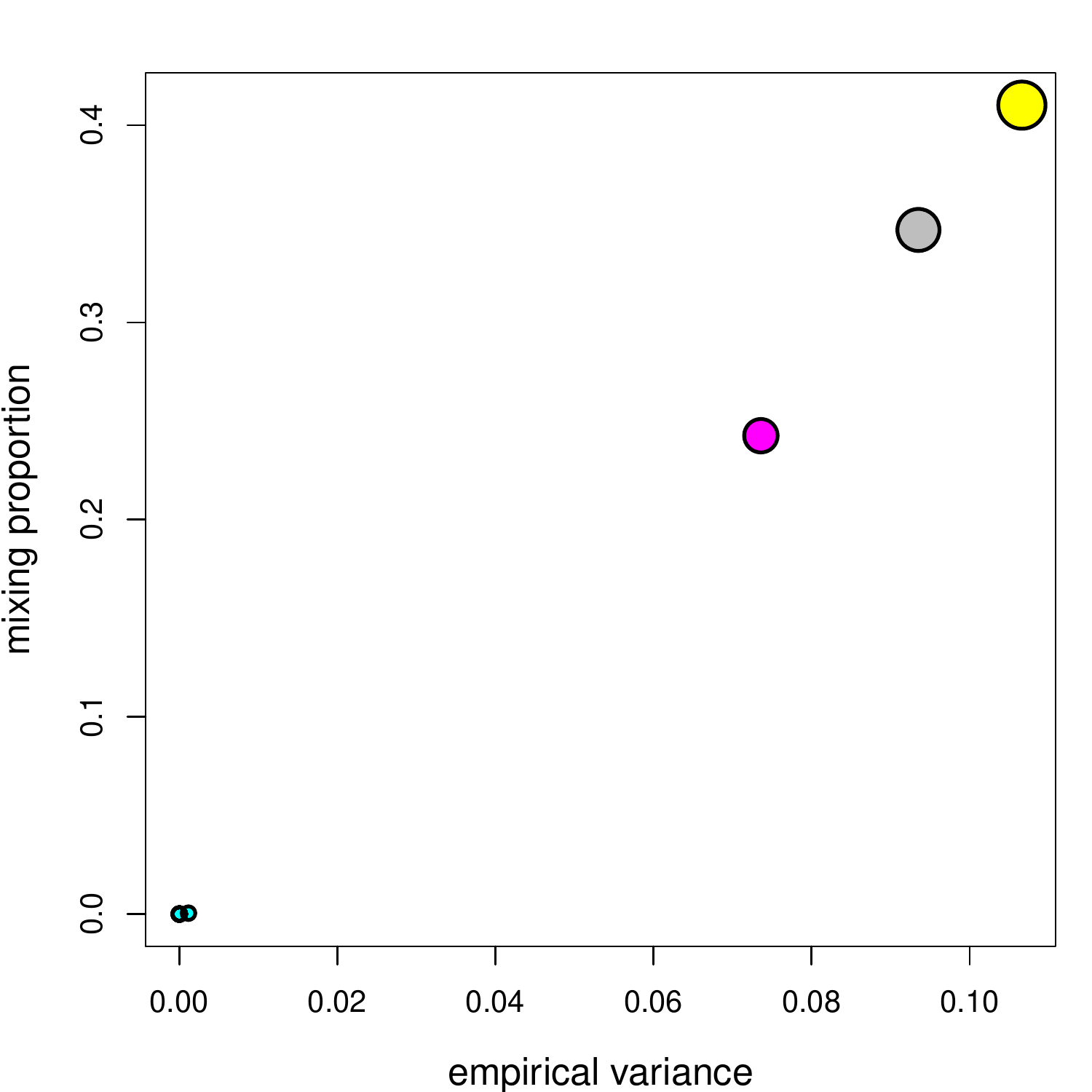}
 \caption{\textbf{Microarray data}. The left panel shows the distribution of the weight log-values assigned to each of the two most relevant dimensions.
 The right panel compares the mixing proportions with the empirical variances of the corresponding latent dimensions.}
 \label{fig:khan_a}
\end{figure}

\begin{figure}[!ht]
 \centering
 \includegraphics[width=0.6\textwidth]{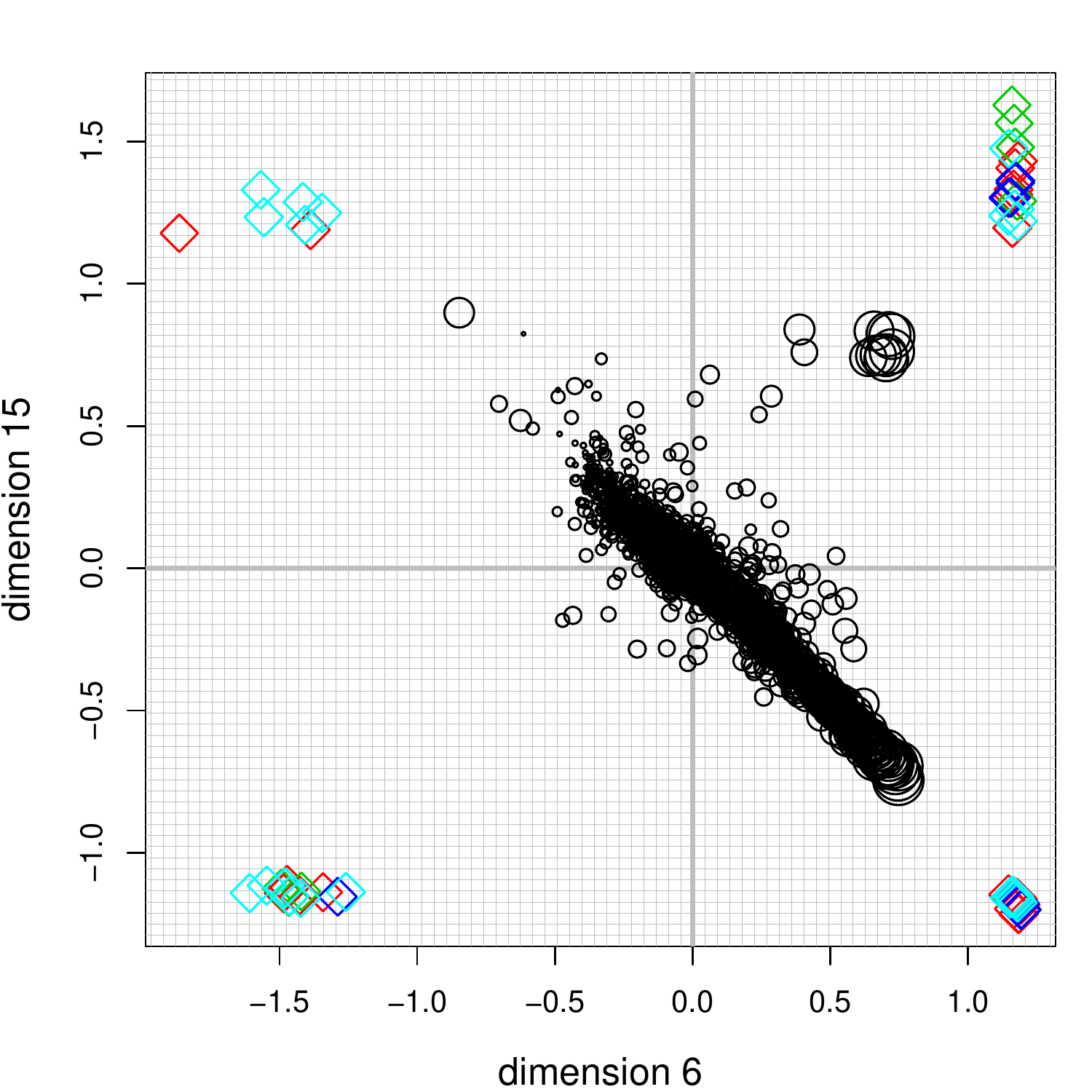}
 \caption{\textbf{Microarray data}. Latent space representation based on the two most relevant latent dimensions.
 The black circles correspond to the genes, whereas the squares correspond to the samples.
 Different square colours indicate the $4$ known tumour classes.
 The sizes of the circles and squares are proportional to the degrees of the corresponding nodes.}
 \label{fig:khan_b}
\end{figure}

The VB algorithm was initialised and run with $K=20$, and converged after $496$ iterations and approximately one hour.
The rounded mixing proportions for the nonempty groups are equal to $\hat{\lambda}_6 = 0.4101$, 
$\hat{\lambda}_{15} = 0.3469$, $\hat{\lambda}_{13} = 0.2425$ and $\hat{\lambda}_{12} = 0.0005$.
As illustrated in Figure \ref{fig:khan_a}, the three most relevant groups all have a relatively small empirical variance,
and they are used to characterise the edges with both large and small weights.
Group $12$ is instead used only for large weights, yet it has a much smaller size and may be considered a group of outliers.
The two largest groups, corresponding to dimensions $6$ and $15$, are used in the latent space representation of Figure \ref{fig:khan_b}.
In both dimensions, the genes are scattered around the origin of the space, whereas the samples are split on both sides of the cloud of genes.
This naturally forms $4$ groups of samples, which do not correspond to the $4$ tumour types included in the metadata.
This is not unexpected, since the separation of the tumour types is known to be a challenging problem, and the SLPM does not involve the optimisation of a clustering criterion on the nodes of the graph.
The results of \textcite{khan2001classification} are of a different nature, 
since the authors use the known tumour classes to train a neural network classifier, and then use this tool to predict the classes on additional test data.

Differently from the hospital data, the genes located close to the origin of the space are those farthest from the samples, hence they have the smallest degrees.
The high degree genes are instead located at the periphery of the cloud, where they are at close distance with at least three of the samples' point clouds.

\section{Conclusions}\label{sec:conclusions}
The SLPM introduced in this paper represents a new research direction for the statistical modelling of networks.
This new framework modifies and extends the LPM approach, 
introducing elements from the literature on blockmodels to enhance the overall flexibility of the proposed model.

Differently from blockmodels and tensor factorisation models, 
the SLPM projects the nodes into a latent space where a weighted transformed Euclidean distance determines the edge values.
This leads to a tool that provides a distance-based representation of the rows and columns of nonnegative interaction matrices,
where the number of latent dimensions is automatically deduced from the data.

The approach proposed prompts a number of new interesting extensions and research possibilities.
Matrix completion and matrix reconstruction frameworks stand out as potential useful applications,
in that the SLPM may represent an effective alternative to NMF methods.
These extensions and the related software are currently being developed.
Furthermore, modifications of the SLPM may also be considered to account for other types of network data, 
such as binary networks, dynamic networks, or multiview networks.

In this paper, a VB framework is proposed to perform inference. 
This method is faster than sampling-based approaches, and it allows one to bypass identifiability issues.
However, the true impact of the approximations introduced and used is not assessed and remains unknown.
Future work may consider more effective strategies that lead to better results in shorter time, 
either using sampling approaches or approximations of a different nature.

\section*{Software}
The \texttt{R} package \texttt{SparseLPM} accompanies this paper and it provides an implementation of the variational algorithm described. 
Parts of the code have been written in \texttt{C++} to reduce overall computing times.
The package is publicly available from \texttt{CRAN} \parencite{rcoreteam}.
All of the computations described in this paper have been performed on a 8-cores (2.2 GHz) Debian machine.

\section*{Acknowledgements}
This research was partially supported through the Vienna Science and Technology Fund (WWTF) Project MA14-031.

% \nocite{*}
\printbibliography

\newpage

\footnotesize
\appendix
\section{Appendix}
Throughout the appendix: $1\leq i \leq M$, $1\leq j \leq N$ and $1\leq k,h \leq K$. 
\subsection{Proof of Proposition \ref{prop:elbo}}\label{app:elbo}
\begin{proof}
We can rewrite the model evidence $p\left( \textbf{X} \right)$ as follows:
\begin{equation}
 p\left( \textbf{X} \right) = \frac{  p\left( \textbf{X} \right) p\left( \textbf{X},\bphi \right)  }{  p\left( \textbf{X},\bphi \right)  }
  = \frac{  p\left( \textbf{X}, \bphi \right)  }{  p\left( \bphi \middle\vert \textbf{X}\right)  }
\end{equation}
Now, we apply a $\log$ transformation, and add and subtract $\log q\left( \bphi \right)$ to the right hand side:
\begin{equation}
 \log p\left( \textbf{X} \right) = \log p\left( \textbf{X},\bphi \right) - \log p\left( \bphi \middle\vert \textbf{X}\right) + \log q\left( \bphi \right) - \log q\left( \bphi \right)
\end{equation}
On the left hand side we have the model $\log$ evidence, which does not depend on $\bphi$. 
Hence, taking the expectations of both sides with respect to $q$ leads to the following:
\begin{equation}
 \log p\left( \textbf{X} \right) = \mathbb{E}_q\left[\log p\left( \textbf{X},\bphi \right)\right] - \mathbb{E}_q\left[\log q\left( \bphi \right)\right] 
 + KL\left[ q\left( \cdot \right) \middle\vert\middle\vert \pi\left( \cdot\middle\vert\textbf{X} \right)\right]
\end{equation}
It is clear that, if we define the variational free energy as follows:
\begin{equation}\label{eq:elbo_2}
  \mathcal{F}\left( \tbphi \right) = \mathbb{E}_q\left[\log p\left( \textbf{X},\bphi \right)\right] - \mathbb{E}_q\left[\log q\left( \bphi \right)\right] 
\end{equation}
minimising the KL divergence is equivalent to maximising $\mathcal{F}$, as $\log p\left( \textbf{X} \right)$ does not depend on $\bphi$.
Also, since the KL divergence is always nonnegative, $\mathcal{F}$ functions as a lower bound for the model evidence.

What remains to be proved is that the definition of the variational free energy given in \eqref{eq:elbo_1} corresponds to the one derived in \eqref{eq:elbo_2}, up to addition by a constant.
The functional $\mathcal{F}$ can be decomposed into a number of terms as follows:
\begin{equation}\label{eq:elbo_3}
 \begin{split}
  \mathcal{F}\left( \tbphi \right) 
  &= \mathbb{E}_q\left[\log p\left( \textbf{X},\bphi \right)\right] - \mathbb{E}_q\left[\log q\left( \bphi \right)\right] \\
  &= \mathbb{E}_q\left[\ell_{\textbf{X}} \left( \mathcal{Z}, \textbf{U}, \textbf{V}\right)\right]
  + \mathbb{E}_q\left[\log \pi\left( \mathcal{Z} \middle\vert \blambda \right)\right] 
  + \mathbb{E}_q\left[\log \pi\left( \blambda \middle\vert \bdelta \right)\right] \\
  &\hspace{1cm}+ \mathbb{E}_q\left[\log \pi\left( \textbf{U} \middle\vert \bgamma \right)\right]
  + \mathbb{E}_q\left[\log \pi\left( \textbf{V} \middle\vert \bgamma \right)\right]
  + \mathbb{E}_q\left[\log \pi\left( \bgamma \middle\vert \textbf{a}, \textbf{b} \right)\right] \\
  &\hspace{1cm}- \mathbb{E}_q\left[\log q\left( \mathcal{Z} \middle\vert \tblambda \right)\right] 
  - \mathbb{E}_q\left[\log q\left( \tblambda \middle\vert \tbdelta\right)\right] \\
  &\hspace{1cm}- \mathbb{E}_q\left[\log q\left( \textbf{U} \middle\vert \tbalpha_{U}, \tbbeta_{U} \right)\right] 
  - \mathbb{E}_q\left[\log q\left( \textbf{V} \middle\vert \tbalpha_{V}, \tbbeta_{V} \right)\right] 
  - \mathbb{E}_q\left[\log q\left( \bgamma \middle\vert \ta, \tb \right)\right] 
 \end{split}
\end{equation}
Before characterising the expectations on the right hand side of \eqref{eq:elbo_3}, we focus our attention of the random variables
\begin{equation}
 \begin{split}
  \Delta_{ijk} &= U_{ik} - V_{jk} \\
  \theta_{ijk} &= \Delta_{ijk}^2 = \left( U_{ik} - V_{jk} \right)^2
 \end{split}
\end{equation}
According to $q$, we know that $\Delta_{ijk} \sim Gaussian\left( \talpha_{Uik} - \talpha_{Vjk}, \tbeta_{Uik} + \tbeta_{Vjk} \right)$ 
and that the variational density on $\theta_{ijk}$ does not have, in general, an explicit form.
However, we do know its expectation and variance:
\begin{equation}
 \begin{split}
  \teta_{ijk} &:= \mathbb{E}\left[ \theta_{ijk} \right] = \mathbb{E}\left[ \Delta_{ijk}^2 \right] 
  = Var_q\left( \Delta_{ijk} \right) + \left( \mathbb{E}\left[ \Delta_{ijk} \right] \right)^2
  = \tbeta_{Uik} + \tbeta_{Vjk} + \left( \talpha_{Uik} - \talpha_{Vjk} \right)^2 \\
  \tzeta_{ijk} &:= Var_q\left( \theta_{ijk} \right) = \mathbb{E}\left[ \Delta_{ijk}^4 \right] - \left( \mathbb{E}\left[ \Delta_{ijk}^2 \right]  \right)^2 \\
  &= \left( \talpha_{Uik} - \talpha_{Vjk} \right)^4 + 6\left( \talpha_{Uik} - \talpha_{Vjk} \right)^2\left( \tbeta_{Uik} + \tbeta_{Vjk} \right) + 3\left( \tbeta_{Uik} + \tbeta_{Vjk} \right)^2 \\
  &\hspace{1cm}- \left( \tbeta_{Uik} + \tbeta_{Vjk} \right)^2 - 2\left( \talpha_{Uik} - \talpha_{Vjk} \right)^2\left( \tbeta_{Uik} + \tbeta_{Vjk} \right) - \left( \talpha_{Uik} - \talpha_{Vjk} \right)^4 \\
  &= 4\left( \talpha_{Uik} - \talpha_{Vjk} \right)^2\left( \tbeta_{Uik} + \tbeta_{Vjk} \right) + 2\left( \tbeta_{Uik} + \tbeta_{Vjk} \right)^2 \\
  &= 2\teta_{ijk}^2 - 2\left( \talpha_{Uik} - \talpha_{Vjk} \right)^4
 \end{split}
\end{equation}
These two quantities play a fundamental role and are used to approximate the expectation of the $\log$-likelihood with respect to $q$.

Now, we evaluate the expectations in \eqref{eq:elbo_3}.
\paragraph{Expectation of the $\log$-likelihood.}
\begin{equation}\label{eq:elbo_4}
\begin{split}
 \mathbb{E}_q\left[\ell_{\textbf{X}} \left( \mathcal{Z}, \textbf{U}, \textbf{V}\right)\right] 
 &= \mathbb{E}_q\left[ \sum_{i,j,k} Z_{ijk}\left\{\log\theta_{ijk} - x_{ij}\theta_{ijk}\right\} \right] \\
 &= \sum_{i,j,k} \mathbb{E}_q\left[Z_{ijk}\right]\left\{\mathbb{E}_q\left[\log\theta_{ijk}\right] - x_{ij}\mathbb{E}_q\left[\theta_{ijk}\right]\right\} \\
 &= \sum_{i,j,k} \tlambda_{ijk}\left\{\mathbb{E}_q\left[\log\theta_{ijk}\right] - x_{ij}\teta_{ijk}\right\} \\
 &\approx \sum_{i,j,k} \tlambda_{ijk}\left\{\psi\left( \frac{\teta_{ijk}^2}{\tzeta_{ijk}} \right) - \log\left( \frac{\teta_{ijk}}{\tzeta_{ijk}} \right) - x_{ij}\teta_{ijk}\right\}
\end{split}
\end{equation}
The approximation of the expectation in the last line of \eqref{eq:elbo_4} comes from the fact that we have replaced the density of $\theta_{ijk}$ 
with a gamma density calibrated to have the same mean and variance.
In fact, neither the true density has an explicit form, nor we can evaluate the expectation of $\log \left(\theta_{ijk}\right)$ exactly.
Once the replacement is made, the expectation of the $\log$ of a gamma variable is readily available.
Other types of approximations may be considered to evaluate this expectation, 
such as Monte Carlo estimators or quadrature methods, however these require a much larger computational cost.
Taylor approximations are more often used (see for example \cite{salter2013variational} and \cite{gollini2016joint}).
The choice of this gamma calibration is purely pragmatic:
we find that, in this particular framework, the approximation proposed leads to much smaller errors than a second order Taylor approximation (not shown).

\paragraph{Expectation of the log prior on $\mathcal{Z}$.}
\begin{equation}
 \begin{split}
  \mathbb{E}_q\left[\log \pi\left( \mathcal{Z} \middle\vert \blambda \right)\right] 
  &= \mathbb{E}_q\left[\sum_{i,j,k} Z_{ijk}\log\lambda_k \right] = \sum_{i,j,k} \mathbb{E}_q\left[ Z_{ijk}\right]\mathbb{E}_q\left[\log\lambda_k \right] \\
  &= \sum_{i,j,k} \tlambda_{ijk}\left[\psi\left( \tdelta_k \right) - \psi\left( \sum_h \tdelta_h \right)\right] \\
 \end{split}
\end{equation}

\paragraph{Expectation of the log prior on $\blambda$.}
\begin{equation}
 \begin{split}
\mathbb{E}_q\left[\log \pi\left( \blambda \middle\vert \bdelta \right)\right] 
&= \log\Gamma\left( \sum_k \delta_k \right) - \sum_k \log\Gamma\left( \delta_k \right) + \sum_k\left( \delta_k - 1 \right)\mathbb{E}_q\left[\log\lambda_k \right] \\
&= const + \sum_k\left( \delta_k - 1 \right)\left[ \psi\left( \tdelta_k \right) - \psi\left( \sum_h \tdelta_h \right)\right] \\
 \end{split}
\end{equation}

\paragraph{Expectation of the log prior on $\textbf{U}$.}
\begin{equation}
 \begin{split}
\mathbb{E}_q\left[\log \pi\left( \textbf{U} \middle\vert \bgamma \right)\right]
&= \sum_{ik} \mathbb{E}_q\left[ - \frac{\log\left( 2\pi \right)}{2} + \frac{\log\gamma_k}{2} - \frac{\gamma_kU_{ik}^2}{2}\right] \\
&= const + \frac{M\sum_{k}\mathbb{E}_q\left[\log\gamma_k\right]}{2} - \sum_{ik}\frac{\mathbb{E}_q\left[\gamma_k\right]\mathbb{E}_q\left[U_{ik}^2\right]}{2} \\
&= const + \frac{M}{2}\sum_{k}\left[ \psi\left( \ta_k \right) - \log\tb_k\right] - \frac{1}{2} \sum_{ik} \frac{\ta_k}{\tb_k}\left( \tbeta_{Uik} + \talpha_{Uik}^2 \right)
\end{split}
\end{equation}
An equivalent result holds for the expectation of the prior on $\textbf{V}$.

\paragraph{Expectation of the log prior on $\bgamma$.}
\begin{equation}
 \begin{split}
\mathbb{E}_q\left[\log \pi\left( \bgamma \middle\vert \textbf{a}, \textbf{b} \right)\right]
&= \sum_k \mathbb{E}_q\left[ a_k\log b_k - \log\Gamma\left( a_k \right) + \left( a_k-1 \right)\log\gamma_k - b_k\gamma_k \right] \\
&= const + \sum_k \left( a_k-1 \right)\mathbb{E}_q\left[\log\gamma_k\right] - \sum_kb_k\mathbb{E}_q\left[\gamma_k\right] \\
&= const + \sum_k \left( a_k-1 \right)\left[ \psi\left( \ta_k \right) - \log\tb_k\right] - \sum_k b_k\frac{\ta_k}{\tb_k}
 \end{split}
\end{equation}

\paragraph{Entropy terms.}
The remaining expectations correspond to the negative entropies of common distributions, whose formulas are well-known.
\begin{equation}
 \begin{split}
\mathbb{E}_q\left[\log q\left( \mathcal{Z} \middle\vert \tblambda \right)\right] &= \sum_{i,j,k} \tlambda_{ijk} \log\tlambda_{ijk} \\
\mathbb{E}_q\left[\log q\left( \tblambda \middle\vert \tbdelta\right)\right] &= 
\log\Gamma\left( \sum_k \tdelta_k \right) + 
\sum_k\left\{ \left( \tdelta_k-1 \right)\left[ \psi\left( \tdelta_k \right) - \psi\left( \sum_h\tdelta_h \right)\right] -\log\Gamma\left( \tdelta_k \right)\right\} \\
\mathbb{E}_q\left[\log q\left( \textbf{U} \middle\vert \tbalpha_{U}, \tbbeta_{U} \right)\right]  &= const - \frac{1}{2}\sum_{ik}\log\left( \tbeta_{Uik} \right) \\
\mathbb{E}_q\left[\log q\left( \bgamma \middle\vert \ta, \tb \right)\right]  
&= \sum_k \left\{ \left( \ta_k-1 \right)\psi\left( \ta_k \right) - \ta_k - \log\Gamma\left( \ta_k \right) + \log\left( \tb_k \right)\right\}
 \end{split}
\end{equation}

The variational free energy in \eqref{eq:elbo_1} is obtained by combining all the formulas and rearranging the terms.
\end{proof}

\subsection{Proof of Proposition \ref{prop:update_lambda}}\label{app:update_lambda}
\begin{proof}
The maximisation of $\mathcal{F}$ with respect to $\left(\tlambda_{ij1},\dots,\tlambda_{ijK}\right)$, 
and under the constraint $\sum_k \tlambda_{ijk} = 1$, is performed using the method of Lagrange multipliers.
Define the following Lagrangian:
\begin{equation}
\begin{split}
 \mathcal{H}_{ij}\left( \tlambda_{ij1},\dots,\tlambda_{ijK},\xi_{ij} \right) &= \mathcal{F}\left( \tbphi \right) + \xi_{ij}\left(\sum_k \tlambda_{ijk} -1\right) \\
 &= const + \sum_{k} \tlambda_{ijk}\left\{C_k  - \log\tlambda_{ijk} + \xi_{ij}\right\} - \xi_{ij}
\end{split}
\end{equation}
where 
\begin{equation}
 C_k = \psi\left( \frac{\teta_{ijk}^2}{\tzeta_{ijk}} \right) - \log\left( \frac{\teta_{ijk}}{\tzeta_{ijk}} \right) - x_{ij}\teta_{ijk}
\end{equation}
The derivatives of $\mathcal{H}_{ij}$ are equal to the following:
\begin{equation}
 \frac{\partial \mathcal{H}_{ij}}{\partial \tlambda_{ijk}} \left( \tlambda_{ij1},\dots,\tlambda_{ijK},\xi_{ij} \right)
 = C_k - \log\tlambda_{ijk} + \xi_{ij} - 1
\end{equation}
Posing the derivative equal to zero and exponentiating gives the following solution:
\begin{equation}
 \tlambda_{ijk}^*= \exp\left\{ C_k + \xi_{ij} - 1 \right\}
\end{equation}
Now, we do the same with the derivative with respect to $\xi_{ij}$:
\begin{equation}
 \frac{\partial \mathcal{H}_{ij}}{\partial \xi_{ij}} \left( \tlambda_{ij1},\dots,\tlambda_{ijK},\xi_{ij} \right)
 = \sum_k \tlambda_{ijk} - 1 = \exp\left\{ \xi_{ij} \right\}\sum_k \exp\left\{ C_k - 1 \right\} - 1
\end{equation}
and obtain:
\begin{equation}
 \xi_{ij} = -\log\left( \sum_k \exp\left\{ C_k - 1 \right\}  \right)
\end{equation}
which leads to \eqref{eq:update_lambda_1}.
A study of the sign of the partial derivatives confirms that this point is a maximum.
\end{proof}

\subsection{Proof of Proposition \ref{prop:update_delta}}\label{app:update_delta}
\begin{proof}
 The first derivative of $\mathcal{F}$ with respect to $\tdelta_k$, for some $k$, is equal to:
 \begin{equation}
 \begin{split}
  \frac{\partial \mathcal{F}}{\partial \tdelta_{k}} \left( \tbphi \right) 
  &= \left[\frac{\partial \psi}{\partial \tdelta_{k}} \left( \tdelta_k \right) - \frac{\partial \psi}{\partial \tdelta_{k}} \left( \sum_h\tdelta_h \right) \right] 
  \left[ \sum_{i,j} \tlambda_{ijk} + \delta_k - \tdelta_k\right] \\
  &\hspace{1cm}- \psi\left( \tdelta_k \right) - \psi\left( \sum_h\tdelta_h \right)
  + \psi\left( \tdelta_k \right) - \psi\left( \sum_h\tdelta_h \right) \\
  &= \left[\frac{\partial \psi}{\partial \tdelta_{k}} \left( \tdelta_k \right) - \frac{\partial \psi}{\partial \tdelta_{k}} \left( \sum_h\tdelta_h \right) \right] 
  \left[ \sum_{i,j} \tlambda_{ijk} + \delta_k - \tdelta_k\right]\\
 \end{split}
\end{equation}
which is equal to zero iff $\tdelta_k = \delta_k + \sum_{i,j} \tlambda_{ijk}$.
A study of the sign of the partial derivatives confirms that this point is a maximum.
\end{proof}

\subsection{Proof of Proposition \ref{prop:update_ab}}\label{app:update_ab}
\begin{proof}
The first derivatives of $\mathcal{F}$ with respect to $\ta_k$ and $\tb_k$, for some $k$, are equal to:
\begin{equation}
\rowcolors{1}{}{}
 \begin{cases}
  \frac{\partial \mathcal{F}}{\partial \ta_k} \left( \tbphi \right)
  &= \frac{\partial \psi}{\partial \ta_k} \left( \ta_k \right)\left( \frac{M+N}{2} + a_k - \ta_k \right) + 1 - \frac{b_k+S_k/2}{\tb_k} \\
  \frac{\partial \mathcal{F}}{\partial \tb_k} \left( \tbphi \right)
  &= \frac{ -\left( a_k + \frac{M+N}{2} \right)  }{\tb_k} + \frac{  \ta_k\frac{S_k}{2} + b_k\ta_k  }{\tb_k^2}
  = \frac{  \ta_k\frac{S_k}{2} + b_k\ta_k  - \tb_k\left( a_k + \frac{M+N}{2} \right)  }{\tb_k^2}
 \end{cases}
\end{equation}
Posing these equal to zero and solving for $\ta_k$ and $\tb_k$ gives \eqref{eq:update_ab_1}.
A study of the sign of the partial derivatives confirms that this point is a maximum.
\end{proof}

\subsection{Proof of Proposition \ref{prop:update_alpha_beta}}\label{app:update_alpha_beta}
\begin{proof}
Let us simplify the notation by denoting $\alpha = \talpha_{Uik}$, $\beta = \tbeta_{Uik}$, 
$\mathcal{F}\left( \alpha, \beta \right) = \mathcal{F}\left( \tbphi \right)$ and $q\left( \alpha,\beta \right) = q_{\textbf{U}}\left( \tbphi \right)$.
To avoid the fact that $\beta$ is constrained to be positive, we derive the parameter updates for its $\log$ and then transform back.
Hence, let us define the function
\begin{equation}
 \mathcal{H}\left( \alpha, \xi \right) = \mathcal{F}\left( \alpha, e^{\xi} \right)
\end{equation}
where we have used the transformation $\beta = \beta\left( \xi \right) = e^{\xi}$. 
Using the chain rule:
\begin{equation}
 \frac{\partial \mathcal{H}}{\partial \xi}\left( \alpha,\xi \right) 
 = \frac{\partial \mathcal{F}}{\partial \beta} \left( \alpha,e^{\xi} \right)    \frac{\partial \beta}{\partial \xi} \left( \xi \right)
 = e^{\xi} \frac{\partial \mathcal{F}}{\partial \beta} \left( \alpha,e^{\xi} \right)
\end{equation}

Thus, a standard gradient ascent update for $\left( \alpha_t, \xi_t \right)$ simply reads as follows:
\begin{equation}
\rowcolors{1}{}{}
 \begin{cases}
  \alpha_{t+1} = \alpha_t + \varepsilon \frac{\partial \mathcal{F}}{\partial \alpha} \left( \alpha_t,e^{\xi_t} \right)\\
  \xi_{t+1} = \xi_t + \varepsilon e^{\xi_t}\frac{\partial \mathcal{F}}{\partial \beta} \left( \alpha_t,e^{\xi_t} \right)
 \end{cases}
\end{equation}

Now, following \textcite{amari1998natural}, we propose instead a natural gradient ascent, which essentially corrects the direction of the gradient to account for the geometry of the space of functions we optimise over.
In practice, the direction of the gradient is corrected by premultiplying it by the inverse Fisher information associated to the likelihood $q\left( \alpha,\xi \right)$.
The $\log$-likelihood and its derivatives read as follow:
\begin{equation}
 \begin{split}
  \log\left[ q\left( \alpha,\xi \right)\right] &= const - \frac{\xi}{2} - \frac{\left(x-\alpha\right)^2}{2e^{\xi}} \\
  \frac{\partial^2 \log\left[ q\left( \alpha,\xi \right)\right]}{\partial \alpha^2} &= -\frac{1}{e^{\xi}} \\
  \frac{\partial^2 \log\left[ q\left( \alpha,\xi \right)\right]}{\partial \alpha\partial \xi} &= - \frac{\left(x-\alpha\right)}{e^{\xi}} \\
  \frac{\partial^2 \log\left[ q\left( \alpha,\xi \right)\right]}{\partial \xi^2} &= - \frac{\left(x-\alpha\right)^2}{2e^{\xi}} \\
 \end{split}
\end{equation}
Hence, the Fisher information and its inverse are equal to the following:
\begin{equation}
\rowcolors{1}{}{}
 \mathcal{I}\left( \alpha,\xi \right) = \left(\begin{matrix}
  \frac{1}{e^{\xi}} & 0 \\
  0 & \frac{1}{2} \\
 \end{matrix}\right) 
 \hspace{2cm}
  \mathcal{I}^{-1}\left( \alpha,\xi \right) = \left(\begin{matrix}
  e^{\xi} & 0 \\
  0 & 2 \\
 \end{matrix}\right) 
\end{equation}
Using these results, we can state the natural gradient ascent update for $\left( \alpha,\xi \right)$, which reads as follows:
\begin{equation}\label{eq:update_alpha_beta_2}
\rowcolors{1}{}{}
 \begin{cases}
  \alpha_{t+1} = \alpha_t + \varepsilon e^{\xi_t} \frac{\partial \mathcal{F}}{\partial \alpha} \left( \alpha_t,e^{\xi_t} \right)\\
  \xi_{t+1} = \xi_t + 2\varepsilon e^{\xi_t} \frac{\partial \mathcal{F}}{\partial \beta} \left( \alpha_t,e^{\xi_t} \right)
 \end{cases}
\end{equation}
Now, we simply take the $\exp$ of the second line, obtaining:
\begin{equation}
\rowcolors{1}{}{}
 \begin{cases}
  \alpha_{t+1} = \alpha_t + \varepsilon e^{\xi_t} \frac{\partial \mathcal{F}}{\partial \alpha} \left( \alpha_t,e^{\xi_t} \right)\\
  e^{\xi_{t+1}} = e^{\xi_t}\exp\left\{ 2\varepsilon e^{\xi_t} \frac{\partial \mathcal{F}}{\partial \beta} \left( \alpha_t,e^{\xi_t} \right)\right\}
 \end{cases}
\end{equation}
which is equivalent to \eqref{eq:update_alpha_beta_1} once the $\log$ transformation is undone:
\begin{equation}
\rowcolors{1}{}{}
 \begin{cases}
  \alpha_{t+1} = \alpha_t + \varepsilon e^{\xi_t} \frac{\partial \mathcal{F}}{\partial \alpha} \left( \alpha_t,e^{\xi_t} \right)\\
  \beta_{t+1} = \beta_t\exp\left\{ 2\varepsilon \beta_t \frac{\partial \mathcal{F}}{\partial \beta} \left( \alpha_t,\beta_t \right)\right\}
 \end{cases}
\end{equation}
Note that the auxiliary variable $\xi$ does not appear in \eqref{eq:update_alpha_beta_1}.
Finally, we shall prove that a small enough $\varepsilon$ leads to a non-decrease of $\mathcal{F}$.
Let us consider again \eqref{eq:update_alpha_beta_2}, and let us define:
\begin{equation}
 \begin{split}
  \Delta_\alpha &= e^{\xi_t} \frac{\partial \mathcal{F}}{\partial \alpha} \left( \alpha_t,e^{\xi_t} \right) \\
  \Delta_\xi &= 2e^{\xi_t} \frac{\partial \mathcal{F}}{\partial \beta} \left( \alpha_t,e^{\xi_t} \right)
 \end{split}
\end{equation}
Using a first order Taylor expansion around $\left( \alpha_t, \xi_t \right)$, there exists a small $\varepsilon > 0$ such that:
\begin{equation}
\rowcolors{1}{}{}
\begin{split}
\mathcal{H}\left( \alpha_t + \varepsilon\Delta_\alpha, \xi_t + \varepsilon\Delta_\xi \right) &\approx
\mathcal{H}\left( \alpha_t, \xi_t \right) + \varepsilon\nabla \mathcal{H}\left( \alpha_t, \xi_t \right)'
 \left[ \begin{matrix}
  \Delta_\alpha \\
  \Delta_\xi \\
 \end{matrix} \right] \\
 &= \mathcal{H}\left( \alpha_t, \xi_t \right) + 
 \varepsilon\left[ \frac{\partial \mathcal{F}}{\partial \alpha} \left( \alpha_t,\xi_t \right), e^{\xi_t}\frac{\partial \mathcal{F}}{\partial \beta} \left( \alpha_t,e^{\xi_t} \right)\right]
 \left[ \begin{matrix}
         e^{\xi_t} \frac{\partial \mathcal{F}}{\partial \alpha} \left( \alpha_t,e^{\xi_t} \right)  \\[6pt]
         2 e^{\xi_t} \frac{\partial \mathcal{F}}{\partial \beta} \left( \alpha_t,e^{\xi_t} \right)
        \end{matrix}
 \right] \\
 &= \mathcal{H}\left( \alpha_t, \xi_t \right) 
 + \varepsilon e^{\xi_t}\left[ \frac{\partial \mathcal{F}}{\partial \alpha} \left( \alpha_t,e^{\xi_t} \right)\right]^2
 + 2\varepsilon e^{2\xi_t} \left[ \frac{\partial \mathcal{F}}{\partial \beta} \left( \alpha_t,e^{\xi_t} \right) \right]^2 \\
 &\geq \mathcal{H}\left( \alpha_t, \xi_t\right)
\end{split}
\end{equation}
or, equivalently: $\mathcal{F}\left( \alpha_{t+1}, \beta_{t+1}\right) \geq \mathcal{F}\left( \alpha_t, \beta_t\right)$.
\end{proof}

\end{document}